\pdfoutput=0
\documentclass[nohyper,notoc]{article} 
\usepackage{color}
\usepackage{epsfig}
\usepackage{bm}% bold math
\usepackage{amsmath}
\usepackage{graphicx}

\def\bit{\begin{itemize}}
\def\eit{\end{itemize}}
\def\ben{\begin{enumerate}}
\def\een{\end{enumerate}}
\def\beq{\begin{equation}}
\def\eeq{\end{equation}}
\def\bea{\begin{eqnarray}}
\def\eea{\end{eqnarray}}
\def\bq{\begin{quote}}
\def\eq{\end{quote}}
\def \lsim{\mathrel{\vcenter
     {\hbox{$<$}\nointerlineskip\hbox{$\sim$}}}}
\def \gsim{\mathrel{\vcenter
     {\hbox{$>$}\nointerlineskip\hbox{$\sim$}}}}
\def\gappeq{\mathrel{\rlap {\raise.5ex\hbox{$>$}}
{\lower.5ex\hbox{$\sim$}}}}
\def\lappeq{\mathrel{\rlap{\raise.5ex\hbox{$<$}}
{\lower.5ex\hbox{$\sim$}}}}

\def\meg{\mu \to e \gamma}

\def\meee{\mu \to e \bar{e} e}
\def\mec{\mu \to e~{\rm conversion}}

\def\a{\alpha}
\def\b{\beta}
\def\g{\gamma}
\def\d{\delta}

\def\m{\mu}

\begin{document}
\vspace*{-1in}
\renewcommand{\thefootnote}{\fnsymbol{footnote}}
\begin{center}
{\Large {\bf 
``Spin-dependent'' $\bm{\mu \to e}$ Conversion on Light Nuclei}}
\vskip 25pt
{\bf    Sacha Davidson $^{1,}$\footnote{E-mail address:
s.davidson@ipnl.in2p3.fr}   Yoshitaka Kuno  $^{2,}$%\footnote{E-mail address:}
 and Albert Saporta$^{1,}$  }
 
\vskip 10pt  
$^1${\it IPNL, CNRS/IN2P3,  4 rue E. Fermi, 69622 Villeurbanne cedex, France; 
Universit\'e Claude Bernard Lyon 1, Villeurbanne;
 Universit\'e de Lyon, F-69622, Lyon, France
}\\
$^2${\it Department of Physics, Osaka University, 
1-1 Machikaneyama, Toyonaka, Osaka 560-0043, Japan}
\vskip 20pt
{\bf Abstract}
\end{center}

\begin{quotation}
  {\noindent\small 
The experimental sensitivity to $\mu \to e$ conversion will  improve by
four or more orders of magnitude in coming years, making it interesting to consider  the ``spin-dependent'' (SD) contribution to the rate.
This process does not
benefit from the  atomic-number-squared enhancement of the spin-independent (SI)
contribution, but probes different operators.
We give details of our recent estimate of the spin dependent rate, expressed
as a function of operator coefficients at the experimental scale. Then
we explore the prospects for distinguishing  coefficients or models
by using different targets,
both  in an EFT perspective, where 
a geometric representation of different targets as vectors
in coefficient space is introduced, 
and also in  three leptoquark models.
 It is found  that comparing the rate on isotopes
with and without spin could  allow to detect
spin dependent coefficients that are  at least a factor of few larger than the
spin independent ones. Distinguishing among  the axial, tensor and pseudoscalar operators that induce the SD rate  would require calculating the nuclear
matrix elements for the second two.
Comparing the SD rate on nuclei with an odd proton  vs odd neutron
could allow to distinguish operators involving $u$ quarks from those involving $d$ quarks; this  is interesting because  the distinction  is difficult to  make for SI operators.

\vskip 10pt
\noindent
}

\end{quotation}

\vskip 20pt  

\setcounter{footnote}{0}
\renewcommand{\thefootnote}{\arabic{footnote}}

%%%%%%%%%%%%%%%%%%%%%%%%%%%%%%%%%%%%%%%%%%%%%%%%%%%%%%%%%%%%%%%%%%%%%%
%% INTRO        %%%%%%%%%%%%%%%%%%%%%%%%%%%%%%%%%%%%%%%%%%%%%%%%%%%%%%
%%%%%%%%%%%%%%%%%%%%%%%%%%%%%%%%%%%%%%%%%%%%%%%%%%%%%%%%%%%%%%%%%%%%%%
%\newpage
\section{Introduction}
\label{intro}

Charged Lepton Flavour Violation (CLFV) is  New Physics that
must exist;  only the rates are unknown.  In this paper, we
consider $\mu \leftrightarrow e$ flavour change, and
assume that it can be parametrised  by contact interactions
involving Standard Model particles.
Flavour change $\mu \leftrightarrow e$ can be probed
in the decays 
$\meg$\cite{TheMEG:2016wtm} and $\meee$\cite{Bellgardt:1987du},
in $\mec$\cite{Bertl:2006up}
 and in  various meson decays such as $K\to \bar{\mu} e$\cite{PDB}.
In $\mec$, 
a beam of $\mu^-$ impinges on a target, where the $\mu$ is
captured by a nucleus,  and can convert to an electron 
while in orbit.  The COMET\cite{COMET} and Mu2e \cite{mu2e}
experiments, currently under construction, plan to improve the 
 sensitivity  by four orders 
of magnitude, reaching a branching ratio  $\sim 10^{-16}$. 
The PRISM/PRIME proposal \cite{PP} aims to  probe $\sim 10^{-18}$.
These exceptional improvements in experimental  sensitivity
motivate our interest in subdominant contributions
to $\mec$.

 Initial analytic estimates of  the $\mec$ rate were performed  by
 Feinberg and Weinberg \cite{FeinWein}, for promising operators and nuclei.
 A wider range of  nuclei were studied numerically by Shanker\cite{Shanker:1979ap},
 and estimates for  many operators and nuclei  can be found in the review \cite{KO}.
Relativistic effects relevant in heavier nuclei were included in \cite{czarM2}.
The matching of CLFV operators constructed with quarks and gluons, onto
operators constructed with nucleons, was performed in \cite{CKOT}.
The current state of the art is the detailed numerical calculations of
Kitano, Koike and Okada (KKO)\cite{KKO}, who studied all  the CLFV nucleon operators that
contribute {\it coherently} to  $\mec$, for nuclei from Helium to Uranium. 
In such processes,  the amplitude for $\mu \rightarrow e$ conversion  on each nucleon
is coherently summed over the whole nucleus.  
 Like ``spin-independent''(SI)  dark matter scattering, the final  rate therefore
 is enhanced by a factor $\sim A^2$, where $A$ is the atomic number of the nucleus.
 However, other conversion processes are possible. For instance, incoherent $\mec$,
 where the  final-state nucleus is in an excited state, has been discussed
 by various people \cite{Kosmas:1993ch,Shanker:1979ap},and is expected
 to be subdominant with respect to the coherent process.

In a previous letter \cite{CDK},  some of us noted that ``spin-dependent''(SD)  $\mec$ can also  occur,  if the target nuclei have spin(as is the case for Aluminium, the target of the upcoming COMET and Mu2e experiments). 
Although this process does not benefit from the $\sim A^2$ enhancement associated to SI rates, it has the interest of being  mediated by different CLFV operators  from the coherent process. 

The aim of this manuscript is to give details of our calculation,
and explore whether the  SD process  could help distinguish models
or operators, should $\mec$ be observed. 
The operators which could induce SD $\mec$ are listed in section
\ref{sec:ops}.
The conversion rate in Aluminium  is estimated in section \ref{sec:rate},
and the extrapolation to other nuclei is discussed in subsection
\ref{ssec:extrapolation}.  The theoretical uncertainties in
our estimates are briefly discussed  in section \ref{sec:uncertainties}.
Section \ref{sec:implications} explores the
consequences of including  the SD contribution
to the $\mec$ rate, both in the perspective
of obtaining constraints  on operator coefficients
from an upper bound on the branching ratio, and
for discriminating models  when  $\mec$
is observed. This section comes in three parts: we
study three leptoquark models which induce SD and SI conversion,
then  consider the same operators but with arbitrary coefficients,
and calculate a covariance matrix. Finally, we allow
all possible  operators with arbitrary coefficients.
We summarise in section
\ref{sec:summ}.

In our previous letter \cite{CDK}, we showed that the SI and SD operator
coefficients mix under  Renormalisation Group(RG) evolution   between the
experimental and weak scales. The effects of this mixing are significant:
the  largest contribution to the $\mec$  rate  from 
an ``SD'' coefficient at the weak scale, would be via
the RG mixing to an SI coefficient (for example, a 
tensor coefficient at the weak scale  induces a
SI contribution to the rate which is $\sim A^2$
larger than the SD contribution).  In this paper, we
focus on operator coefficients at the experimental scale,
only including the RG evolution in the leptoquark
models of section \ref{ssec:LQ}. The
RG evolution of  the operator coefficients
is summarised in Appendix \ref{app:RGEs}.

%%%%%%%%%%%%%%%%%%%%%%%%%%%%%%%%%%%%%%%%%%%%%%%%%%%%%%%%%%%%%%%%%%%%%%
%% SECT  %%%%%%%%%%%%%%%%%%%%%%%%%%%%%%%%%%%%%%%%%%%%%%%%%%%%%%
%%%%%%%%%%%%%%%%%%%%%%%%%%%%%%%%%%%%%%%%%%%%%%%%%%%%%%%%%%%%%%%%%%%%%%

\section{Operators }
\label{sec:ops}

We are interested in contact interactions that
can mediate $\mec$ on nuclei, at a scale
$\mu_N \sim$ 2 GeV.  The focus of this
manuscript is 
the subset of ``spin-dependent''  interactions, but
for completeness,  all 
 QED$\times$QCD invariant operators that 
mediate   $\mu \to e$ conversion on nuclei are  included.
The  relevant operators  in the quark-level   Lagrangian are \cite{KKO,CKOT}:
\bea
\delta {\cal L} & = & - 2\sqrt{2} G_F   \sum_{Y \in L,R} \left[ 
C_{D,Y}  {\cal O}_{D,Y }
 + \frac{1}{m_t} C_{GG,Y}  {\cal O}_{GG,Y }
+ \sum_{q = u,d,s} ~
 \sum_{O'} 
 C_{O',Y}^{qq} {\cal O}_{O',Y}^{qq} 
\right]
+h.c. 
\label{LVAPST}
\eea
where the two-lepton operators are 
\bea
 {\cal O}_{D,Y } = 
m_\mu  (\overline{e}  \sigma^{\a \b} P_Y  \mu ) F_{\a \b}
&&{\cal O}_{GG,Y } = 
 (\overline{e}  P_Y \mu ) G_{\a\b}  G^{\a\b}
\label{opsdefn2l}
\eea
and  $O' \in\{V,A,S,P,T\}$ labels
 2-lepton 2-quark operators
 in a basis where only the  lepton currents are chiral:
\bea
{\cal O}^{ qq}_{V,Y} = 
(\overline{e} \gamma^\a P_Y \mu ) 
(\overline{q} \gamma_\a  q )
&& {\cal O}^{ qq}_{A,Y} 
= (\overline{e} \gamma^\a P_Y \mu ) 
(\overline{q} \gamma_\a \g_5 q )
 \nonumber \\
{\cal O}^{ qq}_{S,Y} = (\overline{e}  P_Y \mu ) 
(\overline{q}  q )
&, &{\cal O}^{ qq}_{P,Y}  = (\overline{e}  P_Y \mu ) 
(\overline{q} \g_5 q ) 
 \nonumber \\
{\cal O}^{ qq}_{T,Y}& =&
 (\overline{e} \sigma^{\a \b} P_Y \mu ) 
(\overline{q} \sigma _{\a \b} q  )
\label{opsdefn}
\eea
with  $\sigma^{\a \b} = \frac{i}{2}[\g^\a,\g^\b]$ and
$P_L = (1-\g_5)/2$. This choice of non-chiral quark currents
is convenient  for matching onto nucleons. However,
often an operator basis  with  chiral quark currents 
is added to the Lagrangian as
$\delta  {\cal L} = -2 \sqrt{2} G_F \sum  C_{O,YX}  {\cal O}_{O,YX}^{qq}$
\cite{KO,megmW,PSI},
where for instance,
$ {\cal O}_{V,YX}^{qq}  \equiv
(\overline{e} \gamma^\a P_Y \mu ) 
(\overline{q} \gamma_\a P_X q )$.
In  this case, the
 coefficients are related as
 (recall that ${\cal O}^{ qq}_{T,LR}$ vanishes---see appendix C of
 \cite{megmW})
\bea
C^{ qq}_{V,Y} = \frac{1}{2}(C^{ qq}_{V,YR} + C^{ qq}_{V,YL})
&&
C^{ qq}_{A,Y} = \frac{1}{2}(C^{ qq}_{V,YR} - C^{ qq}_{V,YL})
 \nonumber \\
C^{ qq}_{S,Y} = \frac{1}{2}(C^{ qq}_{S,YR} + C^{ qq}_{S,YL})
&&
C^{ qq}_{P,Y} = \frac{1}{2}(C^{ qq}_{S,YR} - C^{ qq}_{S,YL})
 \nonumber \\
C^{ qq}_{T,Y} = C^{qq}_{T,YY} ~~~.
&&
\label{coeffs1}
\eea

In eqn (\ref{LVAPST}), the coefficients and operators are evaluated
close to the experimental scale, at $\mu_N \simeq 2$ GeV. The scale is relevant, because
Renormalisation Group running  mixes the tensor and axial
vector operators (that induce
SD $ \mec$) into the scalar and vector operators (who mediate the
SI process)\cite{CDK}\footnote{The analogous mixing of SD WIMP
scattering operators into SI operators was discussed in
\cite{Uli}.}. This is reviewed in Appendix \ref{app:RGEs}.
Throughout the paper, coefficients
without an explicit scale are assumed to be at $\mu_N$.

To compute the rate for $\mec$, 
the operators  containing quarks should be  matched at the  scale $\mu_N$ onto 
CLFV  operators involving  nucleons and mesons.  The relevant  nucleon
operators are  the four-fermion operators of eqn (\ref{opsdefn}) with
$q \to N$ and $N\in \{n,p\}$. As discussed below,  rather than
include mesons in the Lagrangian, we  approximate their effects
by form factors for some nucleon operators and two
additional operators given in eqn (\ref{OD}). So the
nucleon-level Lagrangian will be
\bea
\delta {\cal L} & = & - 2\sqrt{2} G_F   \sum_{Y \in L,R} \left[ 
C_{D,Y}  {\cal O}_{D,Y }
+ \sum_{N = p,n} ~
 \sum_{O''} 
 \widetilde{C}_{O'',Y}^{NN} {\cal O}_{O'',Y}^{NN} 
\right]
+h.c. 
\label{LN}
\eea
where  $O'' \in\{V,A,S,P,T,Der\}$.

At zero momentum transfer ($\vec{P}_f-\vec{P}_i \to 0$),
we  match  onto operators with nucleon currents,
by replacing
\beq
 \bar{q}(x) \Gamma_O q(x) \rightarrow G^{N,q}_O  \bar{N}(x) \Gamma_O N(x)
\label{GONq}
\eeq
such that 
$\langle  N| \bar{q}(x) \Gamma_O q(x)|N \rangle$  =
$G^{N,q}_O \langle  N| \bar{N}(x) \Gamma_O N(x)|N \rangle$=$G^{N,q}_O\overline{u_N}(P_f) \Gamma_O u_N(P_i) e^{-i(P_f-P_i)x}$,
with  $\Gamma_O \in \{ I,\g_5,\g^\a$, $  \g^\b\g_5,\sigma^{\a\b}\}$.
The constants $ G^{N,q}_O$ obtained at zero-recoil
are given in appendix \ref{app:A}, and we will
assume that they are an acceptable approximation
at the  momentum-transfer of  $\mec$, which
is $|\vec{P}_f-\vec{P}_i|^2=m^2_\mu$.

\begin{figure}[h]
\begin{center}
\epsfig{file=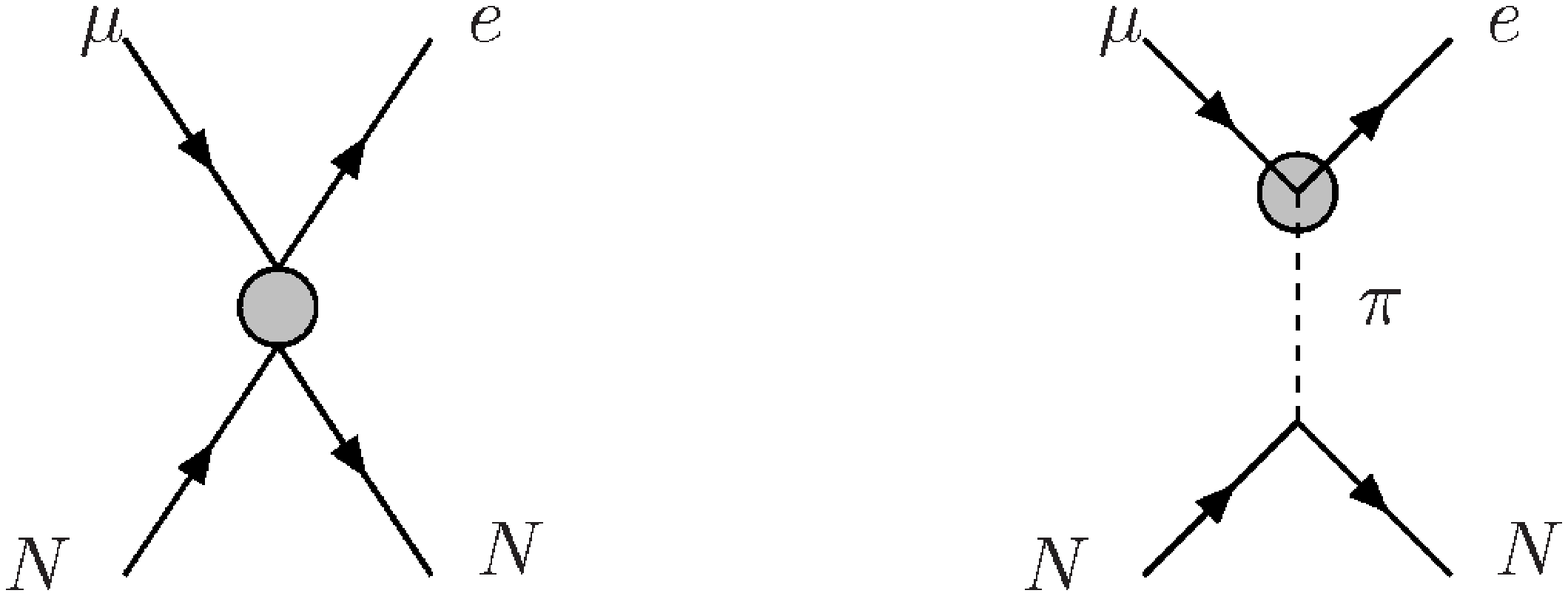, height=4cm,width=12cm}
\end{center}
\caption{ Diagrams contributing to $\mec$ in the presence of
axial and pseudoscalar CLFV operators (represented by the grey blob).
\label{fig:piexchange}
}
\end{figure}

Various mesons are present in the low energy theory at $\mu_N$,
so in principle  the quark operators of eqn (\ref{LVAPST})
should be also matched onto meson operators.   $\chi$PT \cite{Pich:1998xt}
involving nucleons (see {\it e.g.} the
review\cite{Machleidt:2011zz})  would be the appropriate formalism for  this calculation, and has been used  to calculate WIMP
scattering on nuclei \cite{Klos:2013rwa,VCNLO,Hoferichter:2015ipa,Crivellin:2013ipa}, neutrinoless-double-beta-decay \cite{Cirigliano:2017tvr},
and SI $\mec$ \cite{Bartolotta:2017mff}. 
However,  to avoid more notation, here we just give results for
the simple diagrams of interest.  We only
consider  the CLFV decays of pions, because the
effects of heavier mesons would be suppressed by their masses,
and diagrams where a pion is exchanged between
 two nucleons are suppressed by more
propagators, and would require two nucleons
in the initial and final  states\footnote{Such
two-nucleon contributions, which arise at NLO, have been studied in
WIMP scattering \cite{VCNLO,Klos:2013rwa,Hoferichter:2015ipa}, and
recently considered for coherent  $\mec$ in \cite{Bartolotta:2017mff}.}.
Pion decay can
contribute to $\mec$ via the second diagram of figure \ref{fig:piexchange},
in the presence of a pseudoscalar or axial vector quark current.
We  follow the notation of
\cite{Pich:1998xt,Machleidt:2011zz} in matching the axial
vector and pseudoscalar  quark currents
%($O \in A,P$)
onto the
pion, at $P^2 = m_\pi^2$, as
\beq
 \bar{q}(x) \tau^b \gamma^\a \g_5 q(x) \rightarrow    f_\pi i\partial^\a \pi^b(x)
 ~~~,~~
 2m_q \bar{q}(x)\tau^3 \g_5 q(x) \rightarrow    f_\pi m_\pi^2  \pi^0(x)
\eeq
in order to obtain the usual expectation values
$\langle 0| \bar{u}(x) \gamma^\a \g_5 d(x)|\pi^-(P)\rangle$ =  $\sqrt{2} P^\a f_\pi e^{-i P \cdot x}
$, $\langle 0| \bar{u}(x) \gamma^\a \g_5 u(x)|\pi^0(P)\rangle$ =  $ P^\a f_\pi e^{-i P \cdot x}
$,
and  $\langle 0| \bar{u}(x)  \g_5 u(x)|\pi(P)\rangle$ = $f_\pi m_\pi^2 e^{-i  \cdot Px}/2m_u$, where $f_\pi \simeq 92.4$ MeV.

Later in the manuscript, 
the matrix element for $\mu \to e$ conversion on
a nucleon, ${\cal M} (\mu + N(P_i) \to e(k) + N(P_f))$
will be required.  In the case of vector,
scalar  or tensor interactions, it is is straightforward
because conversion proceeds
via a 2-nucleon-2-lepton contact interaction. In the case of axial
vector and pseudoscalar interactions, there is a pion
exchange contribution, as illustrated in figure \ref{fig:piexchange},
so we  give the matrix elements here.  The pion-nucleon interaction
 term in the Lagrangian is taken as $ig_{\pi NN} \overline{N}  \g_5 \vec{\tau} \cdot   \vec{\pi} N $, and  the Goldberger-Treiman relation gives
$g_{\pi pp} \simeq (G_A^{p u}- G_A^{p d}) m_p/f_\pi$. 
 
In the following two equations, $u_N = (u_p,u_n)$
represents a vector of spinors in isospin space.
The matrix element ${\cal M} (\mu + N(P_i) \to e_X(k) + N(P_f))$
mediated by the axial  up quark current, can be written
\cite{EPV,Klos:2013rwa}
\bea
{\Bigg (}\overline{u_N}(P_f)
\frac{[a_0 I + a_1\tau_3]}{2}\g^\a\g_5 u_N(P_i)
% \nonumber \\ &&
+
C^{uu}_{A,X}\frac{ g_{\pi NN}f_\pi q^\a}{|\vec{q^2}| + m_\pi^2}
\overline{u_N}(P_f)[\tau_3]\g_5 u_N(P_i) {\Bigg )}
\overline{u_e}\g_\a P_X u_\mu
\label{MA}
\eea
where $q=(0,-\vec{q}) = P_f-P_i$,   the first term is
written in terms of  iso-scalar and iso-vector contributions
$(a_0 +a_1)/2=  C^{uu}_{A,X} G^{p,u}_A$, $(a_0 -a_1)/2=  C^{uu}_{A,X} G^{n,u}_A$,
whereas the pion contribution is only isovector.

{  In the case of the pseudoscalar operator
${\cal O}^{uu}_{P,Y}$, the pion exchange diagram is non-vanishing
at $|\vec{q}|^2 = 0$, so  at finite momentum transfer, only
the additional contribution $\propto 1/(|\vec{q}|^2 + m_\pi^2) -
1/m_\pi^2$ should be included. This gives}:
\bea
C^{uu}_{P,Y}
{\Bigg (}\overline{u_N}(P_f)
\left[
\begin{array}{cc}
G_P^{p,u} &0\\
0&G_P^{n,u}\\
\end{array}
\right]
\g_5 u_N(P_i)
-
\frac{m_N (G_A^{p,u} - G_A^{n,u})|\vec{q}|^2}{2 m_u(|\vec{q}|^2 + m_\pi^2)}
\overline{u_N}(P_f)[\tau_3] \g_5 u_N(P_i) {\Bigg )}
\overline{u_e} P_Y u_\mu ~~.
\label{MP}
\eea

In summary,  the axial vector and and pseudoscalar quark
operators could equivalently have been matched  at $\mu_N$
to an EFT without pions, but with a $q^2$-dependent
``form factor'' for the  pseudoscalar nucleon operator,
and an additional dimension
seven derivative operator
\beq
{\cal O}^{NN}_{Der,Y} = i(\overline{e} \gamma^\a P_Y \mu ) 
(\overline{N} \stackrel{\leftrightarrow}{\partial_\a} \g_5 N )
\label{OD}
\eeq
such that $i \langle  N(P_f,s')| \bar{N}(x) \stackrel{\leftrightarrow}{\partial_\a} \gamma_5 N(x)|N(P_i,s) \rangle $ =
$ \bar{u}^{s'}_N(P_f) q_\a \gamma_5 u^s_N(P_i)e^{-i(P_f-P_i)\cdot x}$.
In this extended basis, the nucleon coefficients are
\bea
\widetilde{C}_{A,Y}^{NN} & =&  G^{N, u}_A C_{A,Y}^{uu} + G^{N, d}_A C_{A,Y}^{dd} +
G^{N, s}_A C_{A,Y}^{ss}  \label{Ctilde} \\
\widetilde{C}_{Der,Y}^{NN} & =&  \frac{m_\mu m_N}{(m_\mu^2 + m_\pi^2)} {\Big (}G^{N, u}_A- G^{N, d}_A  {\Big )}
 {\Big (}C_{A,Y}^{uu} - C_{A,Y}^{dd} {\Big )}
\nonumber \\
\widetilde{C}_{P,Y}^{NN} & =&  G^{N, u}_P C_{P,Y}^{uu} + G^{N, d}_P C_{P,Y}^{dd} +
G^{N, s}_P C_{P,Y}^{ss} 
-  \left( \frac{C_{P,Y}^{uu}}{2 m_u} - \frac{C_{P,Y}^{dd}}{2 m_d}\right) \frac{m_N (G_A^{N u} - G_A^{N d})m_\mu^2}{(m_\mu^2 + m_\pi^2)}
\nonumber \\
\widetilde{C}_{T,Y}^{NN} & =&  G^{N, u}_T C_{T,Y}^{uu} + G^{N, d}_T C_{T,Y}^{dd} +
G^{N, s}_T C_{T,Y}^{ss} \nonumber \\
\widetilde{C}_{V,Y}^{NN} & =&  G^{N, u}_V C_{V,Y}^{uu} + G^{N, d}_V C_{V,Y}^{dd}  \nonumber 
\eea
where 
$\widetilde{C}_{Der,Y}^{NN}$ was evaluated at $q^2 = -m_\m^2$, and
the  scalar  nucleon
coefficients, to which
contribute also  gluon operator of eqn (\ref{opsdefn2l}),
are given in  \cite{CKOT}.

 To obtain the $\mec$ rate, the expectation values
 of the nucleon operators in the nucleus are required. This
 is discussed in the next section. We were unable
 to find nuclear expectation values of the tensor and
 pseudoscalar operator, so   ${\cal O}_{P,Y}^{NN}$
will be neglected, and the tensor included in
the scalar and axial operators, as described in
eqn  (\ref{redef}).

%%%%%%%%%%%%%%%%%%%%%%%%%%%%%%%%%%%%%%%%%%%%%%%%%%%%%%%%%%%%%%%%%%%%%
%% SECT  %%%%%%%%%%%%%%%%%%%%%%%%%%%%%%%%%%%%%%%%%%%%%%%%%%%%%%
%%%%%%%%%%%%%%%%%%%%%%%%%%%%%%%%%%%%%%%%%%%%%%%%%%%%%%%%%%%%%%%%%%%%%%

\section{Estimating the SD and SI  rate in light nuclei}
\label{sec:rate}

In our previous paper \cite{CDK}, we gave analytic estimates
of the SI and SD conversion rates on Aluminium. 
The aim  of  section  \ref{ssec:CDK} is
to give details of the calculation in the  notation of
 relativistic, second-quantised  Field Theory.
 The results can then be matched onto  the nuclear
 physics calculations of \cite{KKO} (for SI conversion),
 and SD  WIMP scattering \cite{EPV,FHKLX,EngelRTO,Klos:2013rwa} (for SD conversion). 
In subsection \ref{sssec:Comparing}, the  estimates are mapped   onto the
numerical results of  KKO \cite{KKO}, and 
SD conversion in heavier targets is  discussed  in section
\ref{ssec:extrapolation}.

\subsection{Estimating the SD and SI  rate in Aluminium}
\label{ssec:CDK}

  We define
 the   bound state of momentum $P_i$ 
 composed of  an Aluminium  nucleus  and a muon
in the  1$S$ orbital as $\equiv |Al \mu(P_i) \rangle $ . We are  interested in the $S$-matrix element
for  $Al \mu(P_i) \to Al (P_f) + e_X^-(q)$  induced  either
by the dipole operator (which we discuss later), or by  a four-fermion
operator $(\overline{e_X} \Gamma_l \mu ) 
(\overline{N} \Gamma_n N ) $. To be concrete, we consider
the S-matrix element where 
the nucleon $N$ is a proton:
\bea
i 2 \sqrt{2} G_F 
\widetilde{C}_{\Gamma}^{pp}  \langle Al(P_f), e(q,s)|\int  d^4 y 
[\overline{\hat{e_X}}(y) \Gamma_l \hat{\mu}(y) ][ \overline{\hat{p}}(y)
\Gamma_n \hat{p}(y) ] 
| Al\mu  (P_i) \rangle 
\label{S}
\eea
where $s$ is the spin of the electron selected
by the chiral projector $P_X$,   field operators
wear hats, and  $\Gamma_n \in \{ I,\g_5,\g^\a, \g^\b\g_5$, $\sigma^{\a\b}\}$,
 $\Gamma_l \in \{ I,\g^\a,\sigma^{\a\b}\}$.

\subsubsection{four-fermion operators}

\ben
\item A first  step is  
to  write the  motionless
bound state $|Al \mu(0) \rangle $
as
\beq
| Al\m (\vec{P_i} = 0) \rangle =
\sqrt{\frac{2(M_{Al} + m_\m)}{4 M_{Al} m_\m}}
\sum_w
\int \frac{d^3k}{(2\pi)^3}
\widetilde{\psi}_\mu(\vec{k}) \, 
| Al (-\vec{k} ) \rangle \otimes |\m (\vec{k} , w) \rangle 
\eeq
where $w$ is the spin of the muon,  
 the square-root prefactor accounts for 
one vs two-body  normalisation of states 
 in  Lorentz-covariant field theory
 conventions where states are normalised $\propto \sqrt{2E}$ \cite{P+S},
 and $\widetilde{\psi}_\mu(\vec{l})
 = \int d^3 z e^{-i\vec{l} \cdot \vec{z}}\psi_\mu(\vec{z}) $
 is the fourier transform of the
 Schrodinger wavefunction
$\psi_\m (\vec{z})$ for  a muon in a central potential of charge $Z$.

For  $Z\a \ll 1$, the unit-normalised
wavefunction, for either spin state,  can be approximated \cite{Bj+D,Rose,I+Z}
as
\bea
\psi_\mu (r,\theta,\phi) 
\simeq
\frac{[ m \alpha Z]^{3/2}}{\sqrt{\pi } }
 e^{-Z\a mr}  \label{wfnmu} ~~~.
\eea

We  approximate  the outgoing  electron as a  free particle (plane wave), 
which  should be acceptable for an Aluminium target. 
For heavy nuclei, the  Dirac  equation  for the electrons
outgoing in the field of the nucleus should be solved
\cite{czarM2}, allowing to  express
the electron as a superposition of free states.
This approach was followed in KKO \cite{KKO}.

\item{ In the same  non-relativistic  bound state formalism
(see {\it e.g.}, Appendix B of \cite{FHKLX} for more details),
the Aluminium nucleus, of spin $J_A$, can  be written as a bound state
composed of a proton of spin $t$,  with  another state $M_1$  of mass $M_1$
and spin $J_M$ containing $Z-1$
protons and $A-Z$ neutrons:
\bea
\langle Al(P_f),J_A| = 
\sqrt{\frac{2 M_{Al}}{4 M_{1} m_p}}\sum_{t,J_M} \int \frac{d^3 l}{(2\pi)^3} \tilde{f}_p^*(\vec{l},t,J_M,J_A) 
\langle M_1(-\vec{l} + M_1 \vec{v}_f),J_M|
\times \langle p(\vec{l} + m_p \vec{v}_f),t|
\eea
where $\tilde{f}_p(\vec{l},t,J_M,J_A)$ is the fourier transform of the (unknown)
wavefunction of the proton in the potential of $M_1$, and
$P_f =( M_{Al}, M_{Al}\vec{v}_f)$.}

\item   The fermion operators can be expanded  as  \cite{P+S}
\bea
\hat{\mu}(y) = \sum_w \int \frac{d^3p}{(2\pi)^3}\frac{1}{\sqrt{2 E}} {\Big (}
 \hat{a}_p^w u_p^w e^{-ip\cdot y} +  \hat{b}_p^{w \dagger} v_p^w e^{ip\cdot y} 
 {\Big )}
\eea
and act on states as $ \hat{\mu}(y) |\vec{k},w \rangle =
u_k^w e^{-ik\cdot y} |0 \rangle$, where the  spinors  are
normalised as  $u_k^\dagger u_k= 2k_0$. The S-matrix element of eqn
(\ref{S}) can then be evaluated as
\bea
i (2 \pi)^4 \d^4 (P_i - P_f - q) \,
2 \sqrt{2} G_F \widetilde{C}_{\Gamma}^{pp}
\frac{M_{Al}}{m_p \sqrt{2 m_\m}} 
\sum _{p \in Al} \sum_{spins} \int  d^3 x
\psi_\mu (\vec{x})|f_p(\vec{x},J_A,J_M,t)|^2 e^{-i \vec{q} \cdot \vec{x}}
(\overline{u}^s_e \Gamma  u^w_\mu)
(\overline{u}^o_p \Gamma  u^t_p) 
\label{S1}
\eea
where the spinors subscripts are  particle names rather
than momenta,
and $P_i \simeq  (M_{Al} + m_\mu, \vec{P_i}) $,
$P_f \simeq  (M_{Al}, \vec{P_f}) $.
To obtain this approximation, the   states were taken to be
non-relativistic, the wavefunctions expressed
in position space, {  the proton wavefunction was
assumed independent of  the proton spin,} and  the dependence of spinors
on three-momenta was neglected in  many integrals.
 Notice the $M_{Al}/m_p$ enhancement factor that arises
 automatically for both spin-dependent and spin-independent interactions,
 and that the usual $(2 \pi)^4 \d ^4 (P_i - P_f -q)$,  which
 accounts for four-momentum conservation, appears
despite that  there is a spatial integral over the nucleus.
{  In the following, we drop the spin indices
 in the nucleon  distribution in the nucleus $|f_N|^2$.}

\item
The leptonic spinor contraction is independent of $\vec{x}$
and can be factored out of  the spatial integral in eqn (\ref{S1}).
In  light nuclei  such as Aluminium,   the  muon
wavefunction can also be factored out \cite{FeinWein}, 
  because
 the  muon wavefunction decreases on the  scale $\sim 1/(Z\alpha m_\mu)$,
which is larger than the radius of the Aluminium nucleus,
given in \cite{tailleAl} as $\lsim 6 $ fm.
On the other hand, the first zero of the electron
 plane wave (the $e^{-i\vec{q} \cdot \vec{x}}$ of eqn
 (\ref{S1})) would occur 
 at $r\sim \pi/( m_\mu) \sim 6$ fm.

\item  The nucleon spinor contractions, in the non-relativistic limit,
can be written (see eqn (47) of \cite{etalCirelli}):
\bea
\overline{u}^o_N(P_f)  u^t_N(P_i) &\to & 2 m_N \delta^{ot} \nonumber \\
\overline{u}^o_N(P_f) \g_5 u^t_N(P_i) &\to &  2 \vec{q} \cdot \vec{S}_N  \nonumber \\
\overline{u}^o_N(P_f) \g^\a  u^t_N(P_i) &\to & 2 m_N  \delta^{ot} \delta^{\a 0} \nonumber \\
\overline{u}^o_N(P_f) \g^j \g_5 u^t_N(P_i) &\to & 4 m_N  S_N^j   \nonumber \\
\overline{u}^o_N(P_f) \sigma_{ik} u^t_N(P_i) &\to & 4 m_N \epsilon_{ikj} S_N^j   \nonumber \\
\overline{u}^o_N(P_f) \sigma^{0k} u^t_N(P_i) &\to & i q^k
\label{Cirelli}
\eea
where the spin vector of the nucleon is defined as
$2 \vec{S}_N = u_N^\dagger \vec{\Sigma} u_N/{2E_N}$,
and  the rotation
generator $S^{ij} =\frac{i}{4}[\g^i,\g^j] = \frac{1}{2}\epsilon^{ijk} \Sigma^k$.
The momentum transfer $q = P_i - P_f$  has been neglected, except
in the case of the pseudoscalar, where the leading term is ${\cal O}(\vec{q}\cdot \vec{S}_N)$,  and in the case of the tensor, where the there
is a ``spin-independent''   contribution $\propto$ $\vec{q}$.

These spinor identities allow  the tensor interaction involving nucleons
to be absorbed into the scalar and axial vector coefficients. Following
\cite{CDK}, we define
\bea
\widetilde{C}_{S,Y}^{'NN} = \widetilde{C}_{S,Y}^{NN} + 2\frac{m_\mu}{m_N}\widetilde{C}_{T,Y}^{NN} ~~~,~~~ \widetilde{C}_{A,Y}^{'NN}=  \widetilde{C}_{A,Y}^{NN} + 2\widetilde{C}_{T,X}^{NN}
\label{redef}
\eea
where  in both cases the 2 arises from the  two antisymmetric contributions
of the tensor, the unprimed $\widetilde{C}$s are defined in eqn (\ref{Ctilde}), $X,Y \in \{L,R\}$, and  $X\neq Y$ because  only  operators
with  electrons of the same chirality can interfere. { Notice that
there is an error in \cite{CDK}, where is  written 
$\widetilde{C}_{A,Y}^{NN'}=  \widetilde{C}_{A,Y}^{NN} + 2\widetilde{C}_{T,Y}^{NN}$.}

\item It remains to evaluate the expectation
value of the nucleon currents in the nucleus.
\ben
\item
In the case of the scalar or vector operators,
the matrix element of eqn (\ref{S1}) becomes
\beq
{\cal M} = 2\sqrt{2} G_F \widetilde{C}_{S,V}^{pp'}
\frac{2  M_{Al}  }{ \sqrt{2 m_\m}} 
\psi_\mu (0)
\sum_{p \in A}
\int  d^3 x
|f_p(r)|^2\frac{\sin(qr)}{qr}
~ \sum_{s,r}\left\{ \begin{array}{cc} (\overline{u}^s_e   u^r_\mu) & {\rm scalar}\\
(\overline{u}^s_e \g^0  u^r_\mu) & {\rm vector}
\end{array}\right.
\eeq
where  the sum over protons in the
nucleus will give a factor $Z$, we drop the spin indices because
the sum and average   give  one,  and 
assume a spherically symmetric nucleon distribution
$|f_p(r)|^2$ in the nucleus, which allows to replace\footnote{Recall
that  a plane wave can be expanded
on spherical harmonics as $e^{i q z} =
\sum_{\ell = 0}^\infty i^\ell \sqrt{(4\pi)(2\ell +1)} j_\ell(qr) Y_\ell^0(\theta)$,
and $ Y_0^0(\theta) = 1/\sqrt{4\pi}$. }
$ e^{-i \vec{q} \cdot \vec{x}} \to \frac{\sin (qr)}{qr}$.
The ``form factors''  
\beq
F_N(m_\mu) = \int  d^3 x
|f_N(r)|^2\frac{\sin( m_\mu r)}{m_\mu r}
\label{FN}
\eeq
are defined in eqns (29) and (30)  of  \cite{KKO}:
 $F_p(m_\mu) \sim .53$ for Al,  and  $\sim .35$ for Ti.% and  $\lsim 0.2$ for Au.

\item
The expectation value of the   axial   current  in Aluminium ($A=27,Z=13, \vec{J}_{Al} = 5/2$)
was calculated by
 Engel {\it et.al} \cite{EngelRTO}
 and Klos  {\it et.al} \cite{Klos:2013rwa} using   the
shell model.
In the zero-momentum transfer limit, where the spin expectation values
$S_N^A$ are defined by:
\beq
\sum_{N \in A} \int d^3x
|f_N(\vec{x})|^2 
(\overline{u}_N \gamma^k \g_5  u_N) 
=  4 m_{N} S_N^{A} \frac{J_{A}^k}{|J_{A}|} ~~~,
\label{Engel}
\eeq
they obtain  
$  S_n^{Al} =
0.0296$,
  $S_p^{Al} = 0.3430$. ($J_{A}^k$ is a  quantum mechanical
  operator, to be evaluated in the ground state of the nucleus $A$).

At finite momentum transfer, 
references \cite{EngelRTO,Klos:2013rwa} include the
nucleon axial vector operators ${\cal O}^{NN}_{A,X}$  and
 the pion exchange operator ${\cal O}^{NN}_{Der,X}$,
in the combination  induced by axial vector
quark operators. The various terms in the matrix-element-squared
have different spin sums,  so  the  finite momentum transfer correction 
 depends on $\widetilde{C}^{pp'}_{A,X}$ and $\widetilde{C}^{nn'}_{A,X}$,
 and  is quoted  as a multiplicative factor   $S_A(m_\mu)/S_A(0)$
in the rate (see eqn (\ref{BRSD})).
Neglecting $S_n^{Al} \ll S_p^{Al}$, the results of  Engel
{\it et.~al}  for Aluminium give  \cite{EngelRTO}% ($q= m_\mu$)
\bea
S_{Al}(k) &\propto & ( 0.31500480 -1.857857 y +4.86816y^2 - 5.422770y^3)
\label{S(k)}
\eea
where  $y = (m_\mu b/2)^2$ and b =1.73 fm. This gives 
 $S_{Al}(m_\mu)/S_{Al}(0) = 0.29$.

\item  At zero momentum transfer,
the nuclear expectation value of tensor operators ${\cal O}^{NN}_{T,X}$
is proportional  to that of axial vector operators, as
accounted for in eqn (\ref{redef}). However, at finite momentum
transfer, there is  no  pion exchange contribution for the
tensor operator (while pion exchange induces  ${\cal O}^{NN}_{Der,X}$   in
the presence of the  axial vector quark operators),
so the redefinition of eqn  (\ref{redef}) is not valid. Indeed, the
tensor and axial vector operators are distinct
 at finite momentum transfer.

However, we did not find    nuclear calculations
 of SD scattering on Aluminium mediated
 by the tensor operator. We can try to
estimate the error from using the axial  results
for the tensor:    at $q^2 = -m_\mu^2$, the pion exchange
 contribution to the matrix element in eqn (\ref{MA})
is comparable to the  four-fermion contact interaction.
Also, the finite-momentum-transfer suppressions of
the axial and scalar rates on Aluminium are comparable
($S_{Al}(m_\mu)/S_{Al}(0) \simeq |F_N(m_\mu)|^2$),  despite
that one might expect the oscillations of the electron
wavefunction to suppress the SD rate more than the
SI rate, because spin-carrying nucleons are likely
to be at large radii. So we  interpret that
axial matrix element is amplified by a factor $\sim$  2
 at $q^2 = -m_\mu^2$ (due to the pion), and
suppressed by an extra factor $\sim$ 1/2 (as compared
to the scalar matrix element) due to the oscillations of the
electron wavefunction, and  estimate that
the  identification of  eqn (\ref{redef})
could overestimate the tensor contribution to the
branching ratio by
a factor $\sim 2 \to 4$ (depending on
whether the pseudoscalar and
axial matrix elements interfere). 

\item The pseudoscalar operator ${\cal O}_{P,X}^{NN}$  is
proportional to the nucleon spin, is only
present at finite momentum transfer,  and at
$q^2 = -m_\mu^2$,  is enhanced  by a pion
exchange contribution of comparable magnitude. Since
the magnitude of the pseudoscalar spinor contraction
in eqn (\ref{Cirelli}) is  suppressed with respect to
the axial vector by $\sim m_\mu/2m_N$,  its
contribution to the SD branching ratio 
could be  $\sim m_\mu^2/4m_N^2 \times$ the
axial vector contribution.  However, the
identification
$
\widetilde{C}_{A,Y}^{''NN}  =  \widetilde{C}_{A,Y}^{'NN} + \frac{m_\mu}{2 m_N}
\widetilde{C}_{P,X}^{NN}
$
does not work, because  the spin sums  suppress the axial-pseudoscalar
interference term.  A dedicated nuclear calculation would
seem  required for  both the pseudoscalar and tensor operators. 

\een

\item To obtain  the matrix-element-{squared}, the lepton spinor
part can be evaluated by Dirac traces.
Then to perform the nuclear  spin sums in  the SD case, the identity
\bea
\frac{1}{ (2J_{\mu}+1)(2J_{A}+1)}
\sum_{spins} \sum_{k,i}
 \langle J_{\mu}|
\hat{J}^k_\ell
| J'_{e} \rangle
 \langle J'_{e}|
\hat{J}^i_\ell
| J_{\mu} \rangle
 \langle J'_{A}|
\hat{J}^k_A
| J_{A} \rangle
 \langle J_{A}|
\hat{J}^i_A
| J'_{A} \rangle
=\frac{1}{3} J_{\mu}(J_{\mu}+1) J_{A}(J_{A}+1)
\label{BBPSSpinSum}
\eea
can be used.

\item
Finally, the conversion rate is obtained as 
\bea
\Gamma& =&  \frac{1} {2M_{Al}}  \int d\Pi \overline{|{\cal M}|}^2 = 
\frac{m_\mu}{8 M^2_{Al} \pi} \overline{|{\cal M}|}^2  
\nonumber
\eea
where $ \overline{|{\cal M}|}^2$ is averaged over the incident spins, and
$d\Pi$  gives the integration over the final state phase space of the nucleus and electron.
\een

These steps give an analytic estimate for the four-fermion contributions to the
 SI conversion rate on a nucleus of atomic number $A$ and
 charge $Z$:
\bea
\frac{\Gamma_{SI}}{\Gamma_{capt}} &=& 2 B_0
|Z( \widetilde{C}^{'pp}_{S,L} +  \widetilde{C}^{pp}_{V,R} + 2e C_{D,L})F_p(m_\mu) +
(A-Z) (\widetilde{C}^{'nn}_{S,L} + \widetilde{C}^{nn}_{V,R})F_n(m_\mu)|^2
+ \{ L \leftrightarrow R \}~.
\label{BRSI}
\eea
where {  the $F_N$ are defined in eqn (\ref{FN}) and related
to the overlap integrals of KKO in (\ref{FNKKO}), }
the contribution of the dipole operator (estimated in  subsection \ref{dipole})
was also included, and 
$$B_0 = \frac{ G_F^2 m_\mu^5 }{\Gamma_{capt}\pi^2 } (Z\alpha)^3  \simeq \left\{
\begin{array}{ll}
 0.310 & {\rm Al}~  (Z = 13) \\
  0.438 & {\rm Ti} ~ (Z = 22) \\
  \end{array}\right. ~~~, $$
where $\Gamma_{capt}$ is the rate for the Standard Model process of muon
capture ~\cite{KKO,Suzuki:1987jf}.
Similarly, the SD conversion rate  on a nucleus of atomic number $A$, charge
 $Z$ and spin $J_A$ is
\bea
\frac{\Gamma_{SD}}{\Gamma_{capt}}
&=& 8 B_0 \frac{J_{A}+1}{J_{A}}
\, \Big|   S^{A}_p \widetilde{C}_{{A},L}^{' pp}  + 
 S^{A}_n \widetilde{C}_{{A},L}^{' nn} \Big|^2    \  \frac{S_{A} (m_\mu)}{S_{A} (0)} 
+ \{ L \leftrightarrow R \}~.
\label{BRSD}
\eea
where the  spin expectation values $S_N^A$ and
the finite momentum tranfer correction $S_A(k)$ are
given for Aluminium at eqn (\ref{S(k)}).

\subsubsection{the dipole}
\label{dipole}

In the case of the dipole operator of eqn (\ref{opsdefn2l}),
 the S-matrix element can be written
\bea
&&i \frac{2 \sqrt{2} G_F}{\sqrt{ 2m_\mu}} 
{C}_{D,Y}m_\mu   \langle  e(q,s)|\int  d^4 y 
2 (\overline{\hat{e_X}}(y)  \sigma^{0i} \cdot E_i(y) P_Y \hat{\mu}(y) )  
| \mu  (q) \rangle 
\label{Sd} \\
&=& i \frac{2 \sqrt{2} G_F }{\sqrt{ 2m_\mu}} 
{C}_{D,Y} m_\mu  2\pi \delta(E_e - m_\mu)
\int  d^3 y e^{-i\vec{q}\cdot \vec{y}} \psi_\mu(\vec{y}) 
2 (\overline{u_e }  \sigma^{0i} \cdot E_i(y) P_Y {u_\mu}(y) )  
 \\
&\equiv & i 2\pi \delta(E_e - m_\mu) \widetilde{\cal M} ~~~,~~~
 \widetilde{\cal M} = \frac{2 \sqrt{2} G_F }{\sqrt{ 2m_\mu}} 
2 {C}_{D,Y} m_\mu  
\int d \Omega  r^2dr \frac{\sin m_\mu r}{m_\mu r} \psi_\mu(r) 
 (\overline{u_e }  \sigma^{0i}  P_Y  {u_\mu} ) E_i(r)   
\eea
where  the 2  under the integral is to account
for $E_i = F_{0i} = F_{i0}$,  and the magnitude
of the radial  electric field
induced by the nucleus is \cite{KKO}
\beq
E(r)= \frac{Ze}{r^2}\int_0^r r^{'2} |f_p(r')|^2  dr' ~~~.
\eeq
To estimate  the dipole matrix element analytically, we suppose
that  the electric field only contributes at radii within the
first zero of the electron wavefunction  $r_e $,
because outside the rapid oscillation
of the electron wavefunction gives an approximate cancellation
in ${\cal M}$.  The muon wavefunction
is approximately constant at such radii.
The radius of  the Aluminium nucleus
is comparable to $r_e$,  but if we
nonetheless approximate 
the nucleon distribution $ |f_p(r)|^2 $  as a
constant for $r< r_e$, we obtain
\beq
E(r)\simeq \frac{Zer}{3}  |f_p(r)|^2 ~~~,~~~
 \widetilde{\cal M} \simeq  \frac{2 \sqrt{2} G_F }{\sqrt{ 2m_\mu}} 
2 {C}_{D,Y} m_\mu  \psi_\mu(0) 
\int d \Omega  \frac{r^3dr}{3}  |f_p(r)|^2 \frac{\sin m_\mu r}{m_\mu r} 
 (\overline{u_e }  \sigma^{0i}  P_Y  {u_\mu} ) Ze \hat{r}_i
\eeq
where $\hat{r}$ is a radial unit vector.

The ``matrix element'' $\widetilde{\cal M}$ neglects recoil of the
nucleus, so the final state phase space in the rate is only one-body,
and  we  reproduce the analytic estimate of
\cite{KKO} for light nuclei ($D \sim 8 e S^{p}$ given
above eqn (29) of \cite{KKO}): 
\beq
BR_{SI}  =     \overline{|{\cal \widetilde{M}}|}^2 
\frac{m_\mu}{2 \pi}  = \frac{8 G_F^2 m_\mu^5}{ \pi^2\Gamma_{capt} } (\alpha Z)^3
| Z e {C}_{D,Y} F_p(m_\mu)|^2
\eeq
This estimate uses $\int r^3 dr/3 \simeq \int r^2 dr$,
and applies  in the absence of other contributions;  the dipole
coefficient sums with the
 scalar and vector coefficients  in the amplitude, as given
 in eqn (\ref{BRSI}).

\subsubsection{Comparing to KKO}
\label{sssec:Comparing}

This section   compares our estimates
to the more exact formulae of \cite{KKO} (KKO).
Our estimates use a solution of the  Schrodinger
equation for the muon, a plane wave for the electron,
and chiral $\g$-matrices. 
KKO   solve the Dirac
equation in the potential of the nucleus, both for
the electron and muon, use Bjorken and Drell
$\g$-matrix conventions, and give the branching ratio as:
\bea
{\rm BR}(A\mu \to Ae) &=&   \frac{32G_F^2 m_\m^5 }{ \Gamma_{cap}}   
 {\Big [ } \big|     
   \widetilde{C}^{pp}_{V,R} V^{(p)} + \widetilde{C}^{'pp}_{S,L}  S^{(p)}
+ \widetilde{C}^{nn}_{V,R} V^{(n)} + \widetilde{C}^{'nn}_{S,L} S^{(n)} 
+  C_{D,L} {\frac{D}{4}}  
 \big|^2   + \{ L \leftrightarrow R \}~ {\Big ]} 
\label{BRmecKKO}
\eea
where 
$ \Gamma_{cap} $ is the rate for  the  muon to transform
to a neutrino by capture
on the nucleus (see 
~\cite{KKO,Suzuki:1987jf}),  and
the  nucleus- and nucleon-dependent 
``overlap  integrals''  $V_X^{(N)}$, $S_X^{(N)}$,  $D^{(N)}$
correspond to the integral 
over the nucleus of the lepton wavefunctions
and the appropriate  nucleon density (vector, scalar,
electric field for the dipole operator; 
the  definitions and numerical values 
are given in KKO \cite{KKO}). The numerical
coefficient in eqn (\ref{BRmecKKO}) differs from the
result given in KKO, because $4 \widetilde{C}|_{\rm here} = \tilde{g}|_{KKO}$.

Our unit-normalised nuclear density $|f_N(r)|^2$
can be identified with
the similarly normalised  density $\rho_N(r)$ of KKO \cite{KKO}.
Our  Schrodinger  approximation for the muon
wavefunction can be identified to the upper
component (in Bjorken and Drell $\g$ conventions)
of the  Dirac wavefunction obtained by \cite{KKO}. Then the
normalisation conventions of  eqn (5)
and (7) of \cite{KKO} identify 
$$\psi_\mu (r,\theta,\phi) \leftrightarrow \frac{  g_\mu (r)}{\sqrt{4\pi}}~~~. $$

 In the limit of massless electron,  the upper ($g_e$) and
 lower components ($i f_e$) of the electron wave function
  of  \cite{KKO} are comparable. The electron  normalisation
  condition  given in eqn (8) of \cite{KKO} then implies
  that we can identify our  electron plane wave  as 
  $$ if_e= g_e(r) \leftrightarrow \sqrt{2} \frac{\sin  m_\mu r}{ r}
  \leftrightarrow \sqrt{2} m_\mu e^{-i \vec{k}\cdot \vec{r}} ~~.$$

In the approximation where the muon wavefunction
is constant in the nucleus,  the overlap integrals
of \cite{KKO} can be identified to our approximations as
\bea
S^{(p)}, V^{(p)} & \to & \frac{m_\mu |\psi_\mu (0)|}{4\sqrt{\pi}} Z \int d^3x e^{-i \vec{k}\cdot \vec{x}} |f_p|^2 \nonumber\\
S^{(n)},V^{(n)} &\to & \frac{m_\mu^{5/2}(Z\a)^{3/2}}{4\pi} (A-Z) \int d^3x e^{-i \vec{k}\cdot \vec{x}} |f_n|^2 ~~~,
\label{FNKKO}
\eea
as given in eqns (29) - (31) of KKO.

\subsection{Spin-dependent conversion in other light nuclei}
\label{ssec:extrapolation}

In this section we consider how the estimates of the
previous section could be applied to other nuclei.
Recall that light nuclei are interesting for SD  detection,
because the SD rate is relatively suppressed  by $1/A^2$
compared to the SI rate: the ratio
$\Gamma_{SD}/\Gamma_{SI}$ is largest for
light nuclei.

The matrix element  given in eqn (\ref{S1}) for SD $\mec$
contains  the integral of  the axial current over the nucleus,
weighted by the lepton wavefunctions.  
In the case of light nuclei ($Z \lsim 20$),
as discussed in the previous section,
 the muon wavefunction  can be taken constant in the nucleus,
 and the electron can be treated as a plane wave.
 This allows to use the results of  nuclear calculations\cite{EPV}
 of matrix elements for  spin-dependent
 WIMP scattering at finite-momentum-transfer.
 The zero-momentum-transfer matrix elements
(spin expectation values; see eqn (\ref{Engel})),
have been 
calculated for a wide variety of nuclei \cite{Simkovic},
and finite momentum transfer results  also  been  obtained
for some nuclei \cite{Bednyakov:2006ux}. 
 For $\mec$ in heavier nuclei,
 a dedicated   nuclear calculation would
be required  to obtain the  expectation values
of the SD operators  weighted by the lepton wavefunctions.

An interesting light nucleus for SD $\mec$  could be 
Titanium  (Z=22)\footnote{Titanium was  used as a target  by 
 SINDRUMII\cite{Bertl:2006up}, who  set  an upper bound
 $BR (\mu Ti \to e Ti)  < 4.3\times 10^{-12}$.},
because it has isotopes with and without spin,  so
targets  of different isotopic abundances  could
allow to distinguish SD from SI operators. 
Titanium  has  a spin-zero isotope with $A = 48$ and 74\% natural
 abundance \cite{CHEM},  an isotope  with $A = 47, J = 5/2$,
  7.5\%  abundance, and
another isotope with  $A = 49, J = 7/2$, 5.4 \% abundance.
These natural abundances of more than 5 \%
are large enough to make sufficiently-enriched sample targets.

In the Odd Group Model, Engel and Vogel  
\cite{EVdeSim} obtained  spin expectation values
  $S^{Ti,47}_n = 0.21, S^{Ti,47}_p = 0$,  and 
   $S^{Ti,49}_n = 0.29, S^{Ti,49}_p = 0$.
Unfortunately, we were unable to find finite-momentum-transfer
corrections to the spin expectation values in Titanium. 
{  However,  we observe that in Aluminium,   the SI and SD
form factors are comparable: $ 0.28 = |F_p(m_\mu)|^2 \approx
S_{Al}(m_\mu)/S_{Al}(0) = 0.29$.  A similar relation
appears to hold \cite{KKO,Bednyakov:2006ux}  for Florine,
where $ |F_n(m_\mu)|^2 \approx
S_{Fl}(m_\mu)/S_{Fl}(0) \approx .36$. This suggests that
for light nuclei, the  spin-expectation-squared at $|\vec{q}|^2 \neq0$
(that is, $S_{A}(m_\mu)$),  is
similar to the  square of the
 spin-expectation-value at zero momentum transfer, multiplied by
the square of the  SI   $|\vec{q}|^2 \neq 0$ form-factor.
Or taking the square root:}
\beq
\sum_{p \in A} \int d^3x
|f_p(\vec{x})|^2 e^{-i \vec{q} \cdot \vec{x}}
(\overline{u}_p \gamma^k \g_5  u_p) 
\approx \sum_{p \in A}  \int d^3y
|f_p(\vec{y})|^2 %e^{-i \vec{q} \cdot \vec{x}}
(\overline{u}_p \gamma^k \g_5  u_p)
~ \times ~
\int d^3z
|f_p(\vec{z})|^2 e^{-i \vec{q} \cdot \vec{z}}~~~.
\label{approx}
\eeq
So we apply this approximation to Titanium, and estimate
$S_{Ti}(m_\mu)/S_{Ti}(0) \approx 0.12$.

\section{Parametric expansions and uncertainties}
\label{sec:uncertainties}

Once $\mec$ is observed, the aim will be to determine (or constrain)
as many operator coefficients as possible. This would require at least
as many ``independent'' observations as  operators, where observations are
independent if, in spite of uncertainties,
they depend  on a different combination of coefficients.
So the purpose of this section, is to estimate the uncertainties
in relating the conversion rate to operator coefficients. 

The inputs for this calculation, (equivalently,
the theoretical parameters to be extracted from data)
are the coefficients of either the quark operators (see eqn \ref{LVAPST}),
or  of the nucleon
operators (see eqns \ref{Ctilde},\ref{redef}),  in both cases at the
experimental scale $\mu_N$. 
 So  uncertainties associated to the Renormalisation Group evolution from  the
New Physics scale to the experimental scale are not considered. 
In the remainder of this paper, we will sometimes use the quark
operator basis, and sometimes the nucleon basis. 
As discussed in point 1 below,
there are  significant uncertainties in some of the $\{ G_O^{N,q} \}$,
which are  required  to
extract the coefficients of the  quark operators, but can be avoided
by using the nucleon operators. 

\ben

\item 
There are uncertainties  in  some of
the matching coefficients   that relate
quark to hadron  operators (see eqn (\ref{GONq}) and
appendix \ref{app:A}).   The $G_V^{N,q}$ are  from
charge conservation, so should be exact. For   
the axial and scalar coefficients,  the
determinations from data (see eqn (\ref{data})) and
from the lattice(\ref{GAlattice},\ref{GSlattice}) are quoted with smaller
uncertainties than their differences (this
is especially flagrant for the $G_S^{N,q}$,
whose lattice and data values  differ by 30-50\%,
and are both quoted with $\lsim 10\%$ uncertainties).
First, it can be hoped that these discrepancies will
be reduced in the future. Secondly, in some models (or equivalently,
for some choices of coefficients), these factors can be cancelled by
taking ratios. Finally, if we are  only interested in
discriminating SD from SI  contributions
to the rate, this distinction exists at the
nucleon level,  so the matching to quark operators
is not required. 

\item The lepton interactions with nucleons are calculated  at
Leading Order (LO) in $\chi PT$. At NLO, arise  pion
loops as well as processes
with two nucleons in the initial and
final states which exchange a pion that interacts with the leptons.
For the case of WIMP scattering, such NLO contributions
for the  scalar quark operator have been
discussed\cite{Hoferichter:2015ipa,VCNLO,Prezeau:2003sv} and 
reference \cite{VCNLO} estimates them to be a higher order effect
($\lsim 10\%$), provided there are no cancellations
among the LO contributions. The two-nucleon contributions
were also calculated to be unexpectedly small for
WIMP scattering on few-nucleon nuclei \cite{Korber:2017ery}. However,  after this manuscript
was completed,  appeared a study of the $\mec$ rate
mediated  by the scalar and vector interactions \cite{Bartolotta:2017mff}, where
the authors estimate that the NLO effects associated to
pion exchange between two nucleons can reduce
the scalar matrix element  by 20$\to 30\%$(NLO corrections
vanish for the vector). We will account for these
nucleon/$\chi$PT uncertainties by including them in the uncertainties
in the overlap integrals.

\item The $\mec$  matrix element, expressed  as a function
of nucleon operator coefficients, 
 relies on many perturbative expansions, among which
 an expansion in the  finite-momentum-transfer
 $|\vec{q}|^2 = m^2_\m$.   Naively such corrections
 are  ${\cal O}(m_\m^2/m_N^2)$ 
 (so negligible), however  in practise
 there are various effects which are  not  so suppressed. 
First, the finite momentum transfer 
gives a significant suppression of the  matrix element. 
In our analytic approximations,  where
the muon is at rest and the electron  momentum $\vec{k} =\vec{q} $,
this is encoded in the form factors $F_N$
(see eqn (\ref{FN})), which are $\sim .2 \to 0.7$.
KKO  include this  effect  more accurately, 
by solving the Dirac equation for the leptons.
Secondly, finite momentum transfer effects
can change  the nucleon and lepton spinor
algebra. This is discussed for Dark Matter in
\cite{FHKLX,etalCirelli}, and gives the
${\cal O}(m_\m/m_N)$ contribution of the tensor
 to the scalar coefficient given in eqn (\ref{redef}). 
We include this correction, because the
tensor operator at zero momentum transfer  contributes to
the SD matrix element (suppressed by $1/A$),
whereas this $(m_\m/m_N)$-suppressed contribution
gains a relative factor $A$ because it
contributes to the  SI matrix element. The
ratio of these contributions to the  conversion
rate is estimated  in appendix \ref{app:tensor}.
Finally,  pion exchange becomes relevant at
 $|\vec{q}|^2 = m^2_\m$ for the axial vector
 and pseudoscalar operators (see eqns (\ref{MA},\ref{MP})),
 and is included in the nuclear matrix
 elements of \cite{EngelRTO} that we use
 for the axial vector  in Aluminium.
Pion contributions at $|\vec{q}| \neq 0 $
to the SI rate are discussed  above.
% So in summary,
We hope  that  these are the dominant finite-momentum-transfer corrections,
such that any other effects are  negligible ($ < 10\%$) corrections.

\item In our calculation  of the SD matrix element, the
velocity of the initial muon was neglected. This
may seem doubtful, by analogy with
the  extended basis of  WIMP scattering operators
constructed in \cite{FHKLX}, because these authors
expand in both the momentum transfer between the
WIMP and nucleon, and the incoming velocity difference.
However, in our case, the muon velocity is parametrically
smaller: writing
the binding energy of the $1s$ state as $\pi Z \alpha m_\mu \sim m\vec{v}^2$,
gives $|\vec{v}| \sim \sqrt{Z\alpha}$. We neglect any effects related
to this velocity.

\item There  could be nuclear uncertainties in
the SI overlap integrals $S^N, V^N, D$, in addition to the
effects discussed in point 2  above. These were estimated
by \cite{KKO} to be   $\sim$ a few \% in most cases,  $\lsim 10\%$
in the case of  some heavier nuclei. 
\een

Consider first  
the uncertainty on the SI rate, because,  when $\mec$
is observed in a nucleus with spin, the SD conversion rate
can  only be observed, if it is larger than the uncertainty
in the ubiquitous  $A^2$-enhanced SI rate. 
The uncertainty in   $\Gamma_{SI}$, written as a function
of the quark operator coefficients $C_{O,X}^{qq}$, would arise from
the $G_O^{N,q}$,  from  the overlap integrals $S^N, V^N, D$  of \cite{KKO},
and from  NLO contributions in $\chi PT$:
\bea
\frac{\delta \Gamma_{SI} }{ \Gamma_{SI} }(C^{qq}_{O,X})
& \simeq & 2\left( \sum_{X= L,R}\frac{|F_X|
 }{
|F_L|^2 +  |F_R|^2 }{\Big (} C^{qq}_{S,X} S^N \d G^{N,q}_{S} +
 \widetilde{C}^{NN}_{S,X} [\d S^N]_{NLO} {\Big )}  + \frac{\d I_A}{I_A}   \right)
\eea
where
$F_L =   \widetilde{C}^{NN}_{V,L} V^{(N)} + \widetilde{C}^{'NN}_{S,R}  S^{(N)}
+ \widetilde{C}_{D,R} D  $, 
sums on $N \in \{n,p\}$ and $q\in \{ u,d,s\}$ are implicit, the gluon contribution to
the scalar\cite{CKOT} was neglected,
for simplicity 
a common uncertainty $ \frac{\d I_A}{I_A}$ was assigned to the overlap
integrals in nucleus $A$,  except for  the
effect  of neglecting pion
exchange between two nucleons \cite{VCNLO,Bartolotta:2017mff}
(discussed in  point 2 above), which is parametrised as
an uncertainty $[\d S^N]_{NLO}$ in the scalar overlap
integrals.
Expressed this way, the  uncertainty depends on the
quark coefficients  present: 
for $C_{S,X}^{qq} \gg C_{V,X}^{qq}, C_{D,X}$, the
 current discrepancies in the determination
 of the $G_S^{N,q}$ and  $[\d S^N]_{NLO}$   give an ${\cal O}(1)$
 uncertainty on the conversion rate,
 whereas if only the $ C_{V,X}^{qq}$ and $C_ {D,X}$ were present,
the rate uncertainty would come
from the overlap integrals.
The  $G^{N,q}_{S}$ uncertainties can be avoided by
expressing the rate in terms
of the coefficients of the nucleon Lagrangian; if
in addition, $ [\d S^N]_{NLO}/S_N < 10\% $,
then the uncertainty in the SI rate comes from the overlap integrals.
From the KKO discussion,  $2 \frac{\d I_A}{I_A} \lsim 10\%$ in most
cases, $< 20 \%$ in all cases.
In order to be concrete, we assume in the remainder
of this paper, that the uncertainty on the SI rate
expressed in terms of coefficients on nucleons, is
$\lsim 10\%$.
This suggests that the SD rate would need to
be  $\gsim 10-20\%$ of the SI rate, to be observed.

A better sensitivity to the SD rate could be
obtained by using isotopes
with and without spin as targets:
consider  for instance,
 $^{48}$Titanium (without spin),
and  $^{47}$Titanium (with spin), whose
SI matrix elements differ  by one neutron.
Using  the analytic approximation of  eqn (\ref{BRSI}),
the ratio of the SI   conversion rates, for real coefficients and left-handed electrons,
is
\bea
\frac{ \Gamma_{SI} (^{47}Ti) }
{\Gamma_{SI}  (^{48}Ti) }
\simeq
1- 2\frac{ \Big| (\widetilde{C}^{'nn}_{S,L} + \widetilde{C}^{nn}_{V,R})F_n(m_\mu)   \Big|}{| Z( \widetilde{C}^{'pp}_{S,L} +  \widetilde{C}^{pp}_{V,R} + 2e C_{D,L})F_p(m_\mu) +
(A-Z) (\widetilde{C}^{'nn}_{S,L} + \widetilde{C}^{nn}_{V,R})F_n(m_\mu)|}
+... 
\label{cancel}
\eea
where the second term\footnote{Since $^{47}Ti$ and $^{48}Ti$ only
differ by one neutron, there  would be no ${\cal O}(1/A)$   term
if the CLFV operators only involved protons or the dipole.}
is ${\cal O}(1/A)$. The theoretical
uncertainty in this ratio will
arise from the overlap integrals (equivalently, form factors $F_N$),
so  should be of order $\frac{1}{A}\frac {\d I_{Ti}}{I_{Ti}} \lsim 0.002$.
This greatly improves the sensitivity to the
SD rate, although it is unlikely to  allow as  good
a sensitivity to SD  as SI coefficients, because
the SD rate is parametrically suppressed as $1/A^2$
which is  $\lsim \frac{1}{A}\frac {\d I_{Ti}}{I_{Ti}}$.

Some prospects  for distinguishing among SI operators
by using different targets will be discussed in section \ref{ssec:EFT}. 
For this, the various targets need to probe different
combinations of operator coefficients, and this difference
needs to be larger than the theoretical uncertainty. In
section \ref{ssec:EFT}, targets are parametrised
as vectors in coefficient space, whose components
are the overlap integrals (see eqn (\ref{vA})),
and  targets are assumed to probe different combinations of operator
coefficients if the angle between the vectors is
$\gsim  10\% \gsim \frac{\d I_A}{I_A} $.
This estimate can be obtained in the 2-dimensional plane
of the vectors,  where the uncertainty on the angle $\theta$
 of a  point ($I_1 \pm \d I_1 , I_2 \pm \d I_2$) is
\bea
\d \theta \simeq \frac{\d I_i}{I_i}\times  \frac{I_1 I_2}{I_1^2+ I_2^2} 
\label{dt}
 \eea

\section{Implications of  including the SD rate}
\label{sec:implications}

The aim of this section is to explore the implications
of including the SD contribution  to $\mec$.
At first sight, it appears to be of limited interest:
the ratio of SD to SI  rates is
$$
\frac{\Gamma_{SD}}{\Gamma_{SI}} \sim \frac{|C_{SD}|^2}{A^2|C_{SI}|^2}
$$
so for a  SI operator coefficient $C_{SI}$
comparable to $C_{SD}$,  the SD
contribution to the branching ratio is much smaller
than the $\sim$ 10\% theory uncertainty of the  SI contribution,
estimated in the previous section.
Furthermore, as discussed in \cite{CDK}, renormalisation group
running between the New Physics scale and  low energy
mixes the  tensor and axial vector (``SD'') operators
to the scalars and vectors, so even in a  New Physics
model that only induces SD operators, their dominant
contribution to $\mec$ is via the SI operators
 that arise due to RG running. This perspective
 that SD conversion can be ignored is illustrated
 in  section  \ref{ssec:LQ}, where we
 consider three leptoquark models. They give negligeable
 SD branching ratios, but we explore the prospects
 of distinguishing   them by comparing the SI rates in various
 nuclei.

The SD conversion rate is nonetheless interesting,
because it is an independent
observable, that can be  observed by
comparing targets with and without spin. As in
the case of dark matter, it is sensitive to
different operator coefficients (evaluated at
the experimental scale) from the SI process,
so provides complementary information. In
section \ref{ssec:arb}
 we allow  $C_{SD} \gg  C_{SI}$ such that
the SD rate can be observable, and
discuss the  constraints that could
be obtained from upper bounds on
$\mec$. Finally in section \ref{ssec:EFT},
we allow arbitrary coefficients to all
the operators of the nucleon-level Lagrangian, and
explore the  prospects  for
identifying coefficients should
$\mec$ be observed. 

\subsection{Leptoquarks}
\label{ssec:LQ}

We consider three possible leptoquark scenarios,
each containing an
SU(2) singlet leptoquark, 
whose mass  $M \gsim $ few TeV  respects direct search constraints
 \cite{LHCLQ}, and which has only one coupling
to electrons and one to muons. The scenarios are represented
by adding to the Standard Model the following   Lagrangians
\bea
{\cal L}_1 &= &
D^\m S^\dagger D_\m S 
+ M^2 S^\dagger  S
+[\lambda_R^*]_{eu} \overline{e}  u^c S 
+[\lambda_R^*]_{\mu u} \overline{\mu}  u^c S  + h.c. ~~,
\label{L1}
\\
{\cal L}_2 &= &
D^\m S^\dagger D_\m S 
+ M^2 S^\dagger  S  
+[\lambda_L^*]_{\mu d} \overline{\ell}_\mu i \tau_2 q_{L,u}^c S
+[\lambda_R^*]_{eu} \overline{e}  u^c S  + h.c. ~~,
\label{L2} \\
{\cal L}_3 &= &  D^\m \tilde{S}^\dagger D_\m \tilde{S} +
 \widetilde{M}^2 \tilde{S}^\dagger \tilde{S}
+[\tilde{\lambda}^*]_{ed} \overline{e} d^c   \tilde{S}
+[\tilde{\lambda}^*]_{\mu d} \overline{\mu } d^c   \tilde{S} + h.c. ~~.
\label{L3}
\eea
where $D^\mu$ is the appropriate
covariant derivative of QCD and QED. 
At the leptoquark mass scale, we match
these  scenarios   onto  the SM extended by
QED*QCD invariant operators, which mediate $\mec$.
The coefficients and operators  are given in table \ref{tab:LQ}.

\renewcommand{\arraystretch}{1.5}
\begin{table}[htp!]
\begin{center}
\begin{tabular}{|l|l|l|}
\hline
&$~~~~~~~~~~~~~~~~~~~~~~~~~~~~~~~~~$ operators&coefficients at $M$\\
\hline
${\cal L}_1$ &- $\frac{[\lambda_R]^*_{e u} [\lambda_R]_{\mu u}}{M^2}
(\overline{e_R} u^c) (\overline{u^c} \mu_R) $
 = $
 \frac{[\lambda_R]^*_{e u} [\lambda_R]_{\mu u}}{2M^2}
(\overline{e_R}\g^\a \mu_R) (\overline{u} \g_\a P_R u) $& 
$C_{V,R}^{uu} = C_{A,R}^{uu} =
\frac{[\lambda_R]^*_{e u} [\lambda_R]_{\mu u}}{4M^2}$
\\
${\cal L}_2$&-$\frac{[\lambda_R]^*_{eu } [\lambda_L]_{\mu u}}{M^2}
(\overline{e_R} u^c) (\overline{u^c} \mu_L) $
= $
\frac{[\lambda_R]^*_{eu } [\lambda_L]_{\mu u}}{2M^2}
\left(
(\overline{e_R}P_L \mu) (\overline{u}  P_L u)
+\frac{1}{4}
(\overline{e_R}\sigma P_L \mu) (\overline{u} \sigma P_L  u) 
\right)$&$ C_{S,L}^{uu} = 2 C_{T,L}^{uu} =
\frac{[\lambda_R]^*_{e u} [\lambda_L]_{\mu u}}{4M^2}$
\\
${\cal L}_3$ & -$\frac{[\tilde{\lambda}]^*_{ed} [\tilde{\lambda}]_{\mu d}}{\widetilde{M}^2}
(\overline{e_R} d^c) (\overline{d^c} \mu_R)$ = $
 \frac{[\tilde{\lambda}]^*_{e d} [\tilde{\lambda}]_{\mu d}}{2\widetilde{M}^2}
(\overline{e_R}\g^\a \mu_R) (\overline{d} \g_\a P_R d) 
$&$C_{V,R}^{dd} = C_{A,R}^{dd} =
\frac{[\widetilde{\lambda}]^*_{e d} [\widetilde{\lambda}]_{\mu d}}{4\widetilde{M}^2}
$\\
\hline
\end{tabular}
\caption{  Lepton flavour-changing operators
induced in the  leptoquark scenarios of equations
(\ref{L1} -\ref{L3}). The coefficients are given
 at the leptoquark mass scale
$M$, in the basis of section \ref{sec:ops}.
\label{tab:LQ}}
\end{center}
\end{table}

In each scenario, we translate  the coefficients down to the
experimental scale $\mu_{N} = $2 GeV via  an approximate   analytic
solution to  the one-loop
RGEs of QED and QCD \cite{PSI,megmW}:
\bea
 C_I (\mu_N)  &\simeq & C_J(M)\lambda^{a_J}
 \left(
\delta_{JI} - \frac{\alpha_{e}  \widetilde{\Gamma}^e_{JI} }{4\pi} 
\log \frac{M}{\mu_N}  \right) ~~~
\label{oprun1l}
\eea
where
$\lambda = \frac{ \alpha_s(M)}{\alpha_s(\mu_N)} \simeq 1/3  $ for
$M$ = TeV, and 
$I,J$ represent the super- and subscripts which label
operator coefficients. The  
$a_I$ describe  the QCD running
 and are  only non-zero  for scalars and tensors.
 We suppose  five quark flavours for the running,
which gives $a_I = \frac{\Gamma_{II}^s}{2\beta_0} = \{-\frac{12}{23} ,\frac{4}{23} \}$
   for $I = S,T$.
    ${\Gamma}^e$  is  the one-loop  
QED  anomalous dimension matrix,   $\widetilde{\Gamma}^e$
is this matrix with an additional  factor multiplying
the  $TS$ and $ST$ entries ~\cite{Bellucci:1981bs,Buchalla:1989we}
in order to account for the QCD
running:
\bea
\widetilde{\Gamma}^e_{JI} = \Gamma^e_{JI} f_{JI}&,&f_{JI}=\frac{1}{1+a_J - a_I}
\frac{     
\lambda^{a_I - a_J} - \lambda
}{1 - \lambda}   = \left\{
\begin{array}{ll}
\frac{23}{7}\frac{\lambda^{16/23} - \lambda}{1 -\lambda} & JI = ST\\
\frac{23}{39}\frac{\lambda^{-16/23} - \lambda}{1 -\lambda} & JI = TS\\
1&{\rm otherwise}
\end{array}
\right.
\label{fJI}
\eea
 We neglect the RG mixing out of our operator basis, because
 it is small: tensor mixing to the dipole is suppressed by light quark
 masses, and the mixing via the penguin diagram  to vector operators ${\cal O}^{ff}_{V,X}$ is a few \%, and does not generate operators interesting to
 us here. The RG evolution is described in more detail in appendix
\ref{app:RGEs}.

This formalism allows to predict the ratio of
SD to SI contributions to the branching ratio for $\mec$. In
Aluminium, we find for the three scenarios, taking $M = 1 $ TeV:
\bea
{\rm for} ~{\cal L}_1 ~:~~~~~~~\frac{BR_{SD}}{BR_{SI}}
& \sim& 1.5\times 10^{-4} \nonumber \\
{\rm for} ~{\cal L}_2  ~:~~~~~~~\frac{BR_{SD}}{BR_{SI}} & \sim
& 4.4\times  10^{-6}\nonumber\\
{\rm for} ~{\cal L}_3  ~:~~~~~~~\frac{BR_{SD}}{BR_{SI}} &
\sim&  3.2 \times 10^{-5}
\label{BRLQ}
\eea
so we see that the SD rate is smaller than the  current $\sim$10\%
uncertainties on the SI rate, so cannot be  observed in these models.
This is as expected, because the leptoquark model imposes
that the tensor/axial coefficients are
comparable to the scalar/vector coefficients,
so the  SD rate is relatively suppressed
with respect to the SI rate by $1/(AG_S^{N,q})^2$ for tensor coefficients, and
$1/A ^2$ for axial vector coefficients. 

It is interesting to
explore whether the three leptoquark scenarios could be distinguished by
comparing the SI rates in various nuclei. We imagine that $\mec$
has been observed in Aluminium($Z_{Al}$=13, $A_{Al}$ = 27), the target of the
upcoming COMET and Mu2e experiments. We wish to identify
alternative target materials, which could allow to
distinguish our leptoquark scenarios.

A simple distinction between  the
leptoquarks $S$ and  $\widetilde{S}$,  is
that the former couples to $u$ quarks,  and the latter  to $d$ quarks.
To  identify an appropriate target $(A,Z)$,  where the
$\mec$ rates induced by  $S$ and  $\widetilde{S}$
would be significantly different (subject to
the constraint that both reproduced the
Aluminium observations), we consider the double ratio:
\bea
\frac{\left. \frac{\Gamma (Al \mu \to Al e)}{\Gamma ((A,Z) \mu \to (A,Z) e)}\right|_{S}}
{\left. \frac{\Gamma (Al \mu \to Al e)}{\Gamma ((A,Z) \mu \to (A,Z) e)}\right|_{\widetilde{S}}}
\simeq\left(
\frac{2 A_{Al} - Z_{Al}}{ A_{Al} + Z_{Al}}\right)^2 \left(\frac{ A + Z}{2 A - Z}\right)^2
=
\left(\frac{2 V_{Al}^{(p)} +  V_{Al}^{(n)} }{ V_{Al}^{(p)} +  2V_{Al}^{(n)}} \right)^2 \left( \frac{ V_{A}^{(p)} +  2V_{A}^{(n)}}{2 V_{A}^{(p)} +  V_{A}^{(n)} } \right)^2
\label{doubleratio}
\eea
where the operator coefficients cancel because we compare two models
that each induce a single SI operator. 
This ratio should differ from 1 by $\gsim 20$\%, in order to
unambiguously distinguish
the $S$ from $\widetilde{S}$, given the $\sim 10\%$ uncertainties on
the theory calculation.  The first
approximate equality  in eqn (\ref{doubleratio}),
applies for light nuclei, where the conversion rate
can be written as eqn(\ref{BRSI}). The second equality
uses the KKO conversion rate given in
eqn (\ref{BRmecKKO})  in terms of the overlap
integrals $V^{(N)}$, and applies for all nuclei.

\begin{figure}[h]
\begin{center}
\epsfig{file=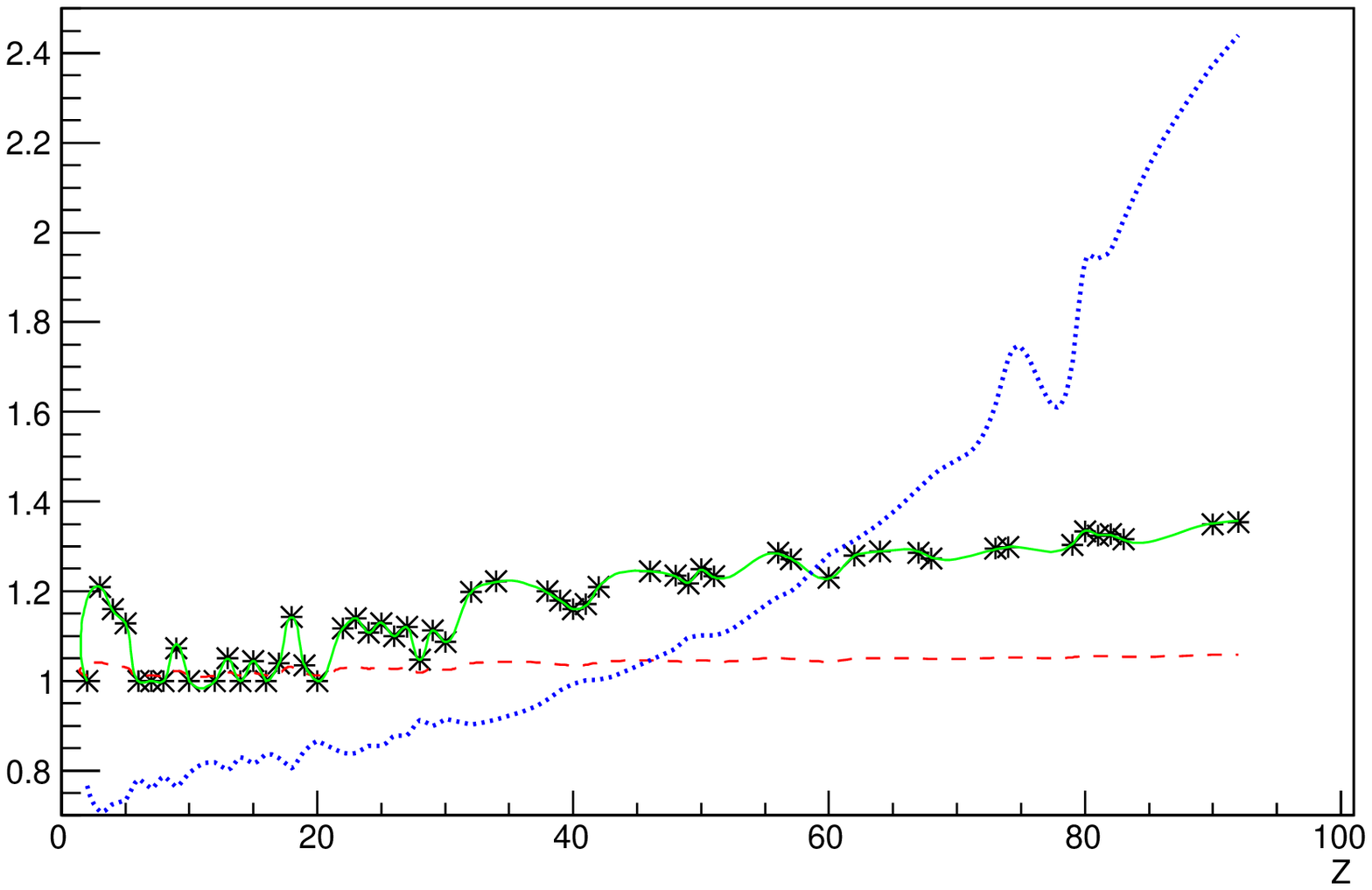, height=8cm,width=12cm}
\end{center}
%\vspace{-10mm}
\caption{This plot illustrates the prospects
for distinguishing SI  operators involving up quarks,
from those involving down quarks, and vector
operators from scalars. The continuous green
[dashed red] line is the
ratio, given in eqn (\ref{doubleratio})  [eqn (\ref{doubleratioS})],
of   $\mec$ rates induced by   ${\cal O}_{V,X}^{uu}$
and ${ \cal O}_{V,X}^{dd}$ [ ${\cal O}_{S,X}^{uu}$
and ${\cal O}_{S,X}^{dd}$], assuming equal coefficients.
The stars on the green line are an analytic approximation.
 The dotted blue line is the ratio, given in eqn (\ref{doubleratioVS}),  of
 $\mec$ rates induced by  ${ \cal O}_{V,X}^{uu}$
and  ${ \cal O}_{S,X}^{uu}$,  with coefficients
selected to give the same rate on Niobium (Z=41).
%ORDI/ROOT/MEC2/plotuvsd.C
\label{fig:uvsd}}
\end{figure}

The continuous green line (with stars) of figure \ref{fig:uvsd} is the ratio of
$\mec$ rates mediated by $S$ and $\widetilde{S}$,
assuming equal operator coefficients. It corresponds to
 the second fraction in the products appearing in eqn (\ref{doubleratio}),
so the double ratio of eqn (\ref{doubleratio}) is simply
obtained by dividing by the ratio for Aluminium. 
The stars are the light nucleus
approximation, the green continous line is the ratio of overlap integrals.
This shows  that the approximation is very similar
to the numerical results of KKO, and  that  a
target with $Z \gsim$ 40  could allow
to distinguish  the first and third leptoquark scenarios.
In the following, we take Niobium (Nb,Z=41,A=93)
as a  $Z \gsim$ 40  target.

It is also interesting 
to explore the
 prospects of distinguishing scalar operators involving
 $u$ vs $d$ quarks. So  we also plot in figure
 \ref{fig:uvsd}, as a dashed red line,
 the  ratio of $\mec$  rates  mediated upstairs
 by ${\cal O}^{uu}_{S,X}$ and downstairs
by ${\cal O}^{dd}_{S,X}$:
\bea
\frac{ \Gamma ((A,Z) \mu \to (A,Z) e) {\Big |}_{{\cal O}^{uu}_{S,X}}}
{ \Gamma ((A,Z) \mu \to (A,Z) e) {\Big |}_{{\cal O}^{dd}_{S,X}}}
=\left(
\frac{G_S^{p,d} S_{A}^{(p)} +  G_S^{n,d}S_{A}^{(n)}}
{G_S^{p,u} S_{A}^{(p)} + G_S^{n,u} S_{A}^{(n)} } \right)^2~~.
\label{doubleratioS}
\eea
For the $G_S^{N,q}$ values given in
 appendix \ref{app:A}, the scalar  ratio is close to one (because
$G_S^{p,q} \simeq G_S^{n,q}$), suggesting that
changing the target in $\mec$ does not help  distinguish   ${\cal O}^{uu}_{S,X}$
from ${\cal O}^{dd}_{S,X}$.

The first and second leptoquark scenarios  respectively
induce scalar and vector operators. As discussed in \cite{KKO,CKOT},
these can be distinguished by comparing the conversion
rates in light and heavy targets.  This is illustrated in
figure \ref{fig:uvsd}, by the blue dotted line, which gives
the double ratio normalised to Niobium
\bea
\frac{\left. \frac{\Gamma (Nb \mu \to Nb e)}{\Gamma ((A,Z) \mu \to (A,Z) e)}\right|_{scalar}}
{\left. \frac{\Gamma (Nb \mu \to Nb e)}{\Gamma ((A,Z) \mu \to (A,Z) e)}\right|_{vector}}
=
\left(
\frac{G_S^{p,u} S_{Nb}^{(p)} +  G_S^{n,u}S_{Nb}^{(n)}}
{G_S^{p,u} S_{A}^{(p)} + G_S^{n,u} S_{A}^{(n)} } \right)^2
\left( \frac{2 V_{A}^{(p)} +  V_{A}^{(n)}}{2 V_{Nb}^{(p)} +  V_{Nb}^{(n)} } \right)^2
~~~.
\label{doubleratioVS}
\eea
 We see that
measuring the  $\mec$ rate on  Aluminium, some intermediate
target around $Z\sim 40$ and on a heavy nucleus like lead or gold (Z = 79),
could distinguish  the three leptoquark scenarios, that is
a vector operator involving $d$s, vs vector operator involving $u$s,
vs a scalar operator involving $u$s. 

\subsection{ Bounds on arbitrary coefficients of four operators}
\label{ssec:arb}

This section considers the  operators   induced by
the second and third leptoquark
models (see equations (\ref{L2}),(\ref{L3}))
which are
added simultaneously to the Lagrangian 
 with arbitrary coefficients:
\bea
{\cal L}_{EFT} &=&
C^{uu}_{S,L}
{\cal O}^{uu}_{S,L}
+C^{uu}_{T,L}
{\cal O}^{uu}_{T,L}
+C^{dd}_{V,R}
{\cal O}^{dd}_{V,R}
+C^{dd}_{A,R}
{\cal O}^{dd}_{A,R}
+ h.c.
\label{LEFT}
\eea
This is clearly an incomplete basis (the complete basis
of dimension six operators at $\mu_N$ is given in eqns
(\ref{LVAPST},\ref{opsdefn})); however, it is sufficient for
our purpose\footnote{In a later publication,
we may try to constrain operator coefficients and
count ``flat directions'', for which a complete basis would be required.}, which is to  explore which constraints can be obtained on
quark-level operators from the non-observation of
 $\mec$ in targets with and without spin.

We suppose that $\mec$ has not been
observed on Aluminium, Titanium (enriched
in isotopes with spin) and Lead targets.
{ These targets are chosen because heavy and
light targets have different sensitivities to
vector and scalar coefficients, and because
the spin of Titanium and Aluminium is respectively
associated to an odd neutron and an odd proton.}
In order to check that  upper bounds on these
branching ratios can constrain all the operator
coefficients  which we consider, 
we set the branching ratios to
zero, and check that this forces 
the coefficients to vanish.

Setting the SD conversion rates in Titanium  and
Aluminium to zero gives two equations:
\bea
0 &\simeq& C^{dd}_{A,R}\left(G_A^{p,d} + \frac{S_n^{Al}}{S_p^{Al}}G_A^{n,d} \right)
+ 2 C^{uu}_{T,L}\left(G_T^{p,u} + \frac{S_n^{Al}}{S_p^{Al}}G_T^{n,u} \right)
 \\
0 &\simeq& C^{dd}_{A,R} G_A^{n,d} 
+ 2 C^{uu}_{T,L} G_T^{n,u} 
\label{distinct}
\eea
where $\frac{S_n^{Al}}{S_p^{Al}} \simeq 0.1$ is the ratio of
spin expectation values in Aluminium. These equations have
solutions 
$$  2 C^{dd}_{A,R}  \simeq C^{uu}_{T,L} 
~~~,~~~  C^{dd}_{A,R}  \simeq 2 C^{uu}_{T,L}
$$
so even allowing for a 10\% theory uncertainty on
the coefficients, it is clear that the only solution
is for both coefficients to vanish. This is  because
the spin of Titanium isotopes  arises from the
odd number of neutrons, whereas in   Aluminium
the spin is from the odd proton, so these two   conversion
rates probe  the SD coefficients $\widetilde{C}^{'NN}_{A,X}$
for both  neutrons  and protons.
Then, since the matching coefficients $G^{Nu}_{A,X}$
and $G^{Nd}_{A,X}$  (equivalently
 $G^{Nu}_{T,X}$
and $G^{Nd}_{T,X}$) 
are of opposite sign and different
magnitude,   $C^{uu}_{A,X} + 2 C^{uu}_{T,X}$
and $C^{dd}_{A,X} + 2 C^{dd}_{T,X}$ can be distinguished. 

It is straightforward to check that setting the
SI rates on Al, Ti and Pb to zero, forces
$C^{dd}_{V,R} , C^{uu}_{S,L} \to 0$. 

 A more informative way to present the constraints
 on coefficients arising from the experimental bounds
 is to give the covariance matrix. We suppose
 an upper bound of $BR$ (for instance, $10^{-14}$) on
 the SI branching ratios on Pb and Al, and
 on the SD branching ratios on Al and Ti.
The tensor operator gives  comparable contributions
to both SI and SD processes (see Appendix \ref{app:tensor}),
so the $4\times 4$ covariance matrix does not split into $2\times 2$
subblocks. Nonetheless, it is interesting to give the
covariance matrices for different cases, in order to see
the variation of the bounds, when different 
 theoretical information is included. 

First,  the tensor contribution to the SI rates  is neglected,
in which case the  covariance matrices for $(C^{dd}_{V,R} , C^{uu}_{S,L})$
 and $( C^{uu}_{T,L}, C^{dd}_{A,R} )$ are:
 \bea
BR \left[
\begin{array}{cc}
0.012 & -.0028\\
 -.0028 & .0007
\end{array}
\right]~~~,~~~
BR \left[
\begin{array}{cc}
 9.1 & 20\\
20 &73.6
\end{array}
\right] ~~~.
\label{22}
\eea
So, for instance,  $|C^{uu}_{S,L}|$ is excluded
above $\sqrt{0.0007\times BR}$, and
 $|C^{dd}_{A,R}| < \sqrt{73.6 \times BR}$.

If now the SD  rates are neglected, but the tensor contribution
to SI is included, then  the  covariance matrix for 
$(C^{dd}_{V,R} , C^{uu}_{S,L},  C^{uu}_{T,L})$ is 
 \bea
BR \left[
\begin{array}{ccc}
0.47 & -.24&23\\
 -.24 & .13&-14\\
 23&  -14&  1 400 
\end{array}
\right] ~~~,
\label{33}
\eea
which  shows that the exclusions become weaker
due to  potential cancellations between a
large  $C^{uu}_{T,L}$ and the  vector or scalar
coefficients. 
Finally  the full covariance matrix arising from
imposing $BR_{SI}(\mu Pb \to e Pb) \leq 10^{-14}$,
$BR_{SI}(\mu Ti \to e Ti) \leq 10^{-14}$,
$BR_{SD}(\mu~ ^{47}Ti \to e~ ^{47}Ti) \leq 10^{-14}$,
$BR_{SI}(\mu Al \to e Al) \leq 10^{-14}$, and 
$BR_{SD}(\mu Al \to e Al ) \leq 10^{-14}$,
for
the  coefficients $(C^{dd}_{V,R} , C^{uu}_{S,L},  C^{uu}_{T,L}, C^{dd}_{A,R} )$, is 
\begin{align}
BR
\left[
\begin{array}{cccc} 
 0.010 & -0.0029  & 0.12 &0.26  \\
-0.0029 & 0.0011   & -0.078  &-0.17 \\
0.12    &-0.078 & 9.0  &19.6 \\
0.26 &-0.17 &19.6 &73\\
\end{array}
\right] ~~~.
\end{align}
Comparing  to  the bounds of eqn (\ref{22}),  shows that
the tensor contribution to the SI rate is of
little importance,  provided the   SD bounds are included. 
However,   if the  SD bounds are neglected,
including the tensor in the SI rate significantly
weakens the constraints, as can be seen in eqn (\ref{33}).
{We  also checked 
that including   $BR_{SI}(\mu Au \to e Au) \leq 10^{-14}$ 
only changes a few entries by about 25\%, as expected, because
 Al, Ti and Pb were chosen as targets for their discriminating power.}

\subsection{Reconstructing nucleon coefficients}
\label{ssec:EFT}

We now suppose that $\mec$ is observed in Aluminium, 
where there can  be  SI and SD contributions to the rate,
and that the New Physics is described by the
nucleon-level Lagrangian of eqn (\ref{LN}) with
arbitrary operator coefficients.
It is interesting to consider
which  subsequent  targets, in what order,
would be required  to distinguish the SD and SI contributions,
and then to discriminate among the SI operators?

We first introduce a geometric representation of  models and targets,
which allows to visualise the ability of various targets to discriminate
among models.
A New Physics scenario  can be represented as a  two 5-dimensional vectors,
each composed of SI 
 coefficients which   interfere 
$\vec{C}_X \equiv ({C}_{D,X}$, $\widetilde{C}_{S,X}^{'pp}$,
$\widetilde{C}_{V,Y}^{pp}$,$\widetilde{C}_{S,X}^{'nn}$,$\widetilde{C}_{V,Y}^{nn})$,
and  two two-component vectors of SD coefficients
$(\widetilde{C}_{A,X}^{'nn}$,$\widetilde{C}_{A,X}^{'pp}).$
For simplicity, we focus on $X=L$, and drop this electron chirality subscript.
Then we focus on discriminating among SI operators, because
the spin of target nuclei is  usually associated   to  either an
unpaired $n$ or $p$, giving   an order of magnitude  better sensitivity
to the coefficient corresponding to the unpaired nucleon (see,
{\it e.g.}  the spin expectation values given after eqn  (\ref{Engel})).
This means that discriminating $\widetilde{C}_{A,X}^{'nn}$
vs $\widetilde{C}_{A,X}^{'pp}$ should be a straightforward
matter of using targets with an unpaired $p$ and $n$.

\begin{figure}[ht]
\begin{center}
\epsfig{file=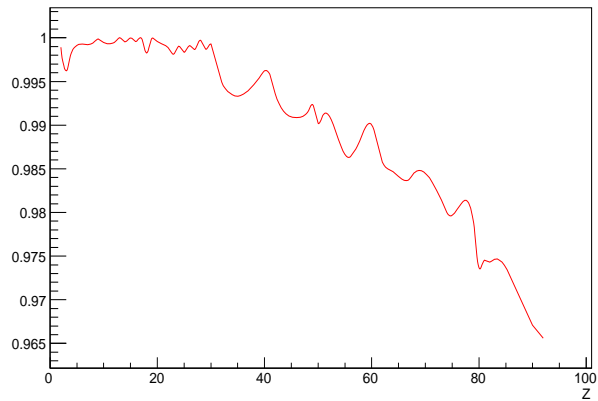, height=6cm,width=9cm}
\end{center}
%\vspace{-10mm}
\caption{ A representation of the discriminating power
of a target (labelled by $Z$), with respect to Aluminium.
On the vertical axis is
the invariant measure, given in eqn (\ref{ipdiff}), of the misalignment
in coefficient space of the target with respect to 
Aluminium.
%ORDI/ROOT/MEC2/plot4.C 
\label{fig:AlAip1}}
\end{figure}

For the spin independent process,
a target nucleus $(Z,A)$ can be envisaged as a vector
\beq
\vec{v}_{(Z,A)}  = (D_{(Z,A)}, S_{(Z,A)}^{(p)}, V_{(Z,A)}^{(p)},S_{(Z,A)}^{(n)}, V_{(Z,A)}^{(p)})
\label{vA}
\eeq
in
the five-dimensional coefficient space,
whose components are the appropriate  overlap integrals.
(In the following, the vectors and components are indiscriminately labelled
by $A$ or $Z$ because  we use the overlap integrals of KKO,
obtained for  a single  abundant isotope.)
The matrix element for
$\mec$ on target $A$,  mediated by a combination of coefficients
$\vec{C}$,
 is proportional to $\vec{C} \cdot \vec{v}_A $,
 and target  nucleus $A$  allows to probe
 coefficients in the direction $  \vec{v}_A $.
If we define the  unit-normalised
 $\hat{e}_A = \vec{v}_A /|\vec{v}_A | $,
then  target $A$ probes the same
 combination of coefficients as Aluminium
 if $\hat{e}_A$ is parrallel to
 $\hat{e}_{Al}$, and 
the difference
\beq
1-\hat{e}_A \cdot \hat{e}_{Al}  \approx \frac{\theta^2}{2}
\label{ipdiff}
\eeq
gives an invariant measure of  whether the target $A$   has
sensitivity to an orthogonal direction
in coefficient space. In eqn (\ref{ipdiff}),
$\theta$ is the angle between $\hat{e}_A $ and $ \hat{e}_{Al}$.
Figure \ref{fig:AlAip1} gives
$\hat{e}_A \cdot \hat{e}_{Al}$
as a function of $Z$.
From  eqn (\ref{dt}),
the uncertainty in the direction of $\hat{e}_A$ is $\lsim 10\%$,
so target $A$ is indistinguishable from Aluminium for
$\hat{e}_A \cdot \hat{e}_{Al}  \gsim 0.995$, or $Z \lsim 25-30$.

\begin{figure}[ht]
\begin{center}
\epsfig{file=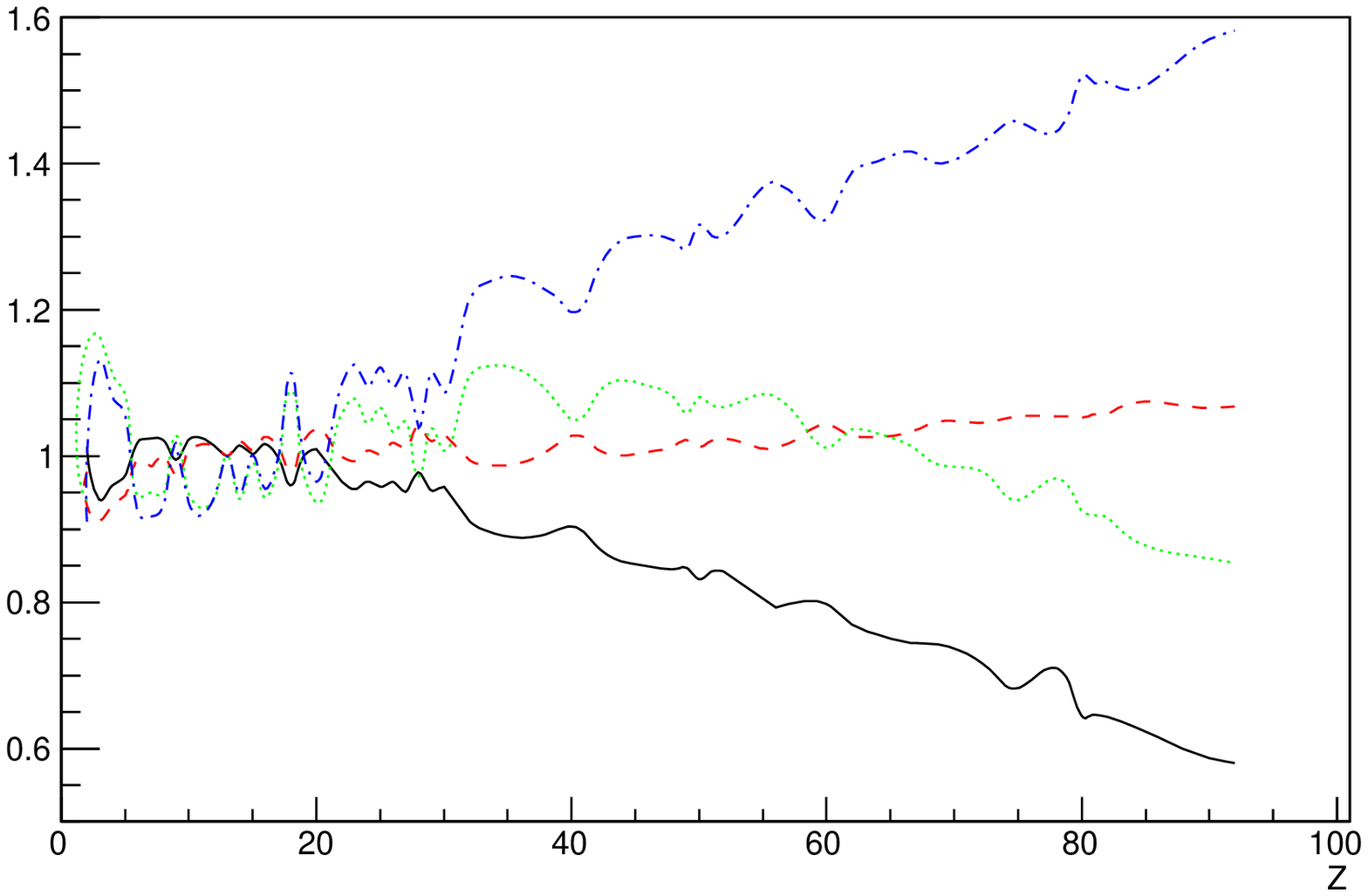, height=6cm,width=9cm}
\end{center}
%\vspace{-10mm}
\caption{ An  operator-dependent measure of the
discriminating power of a targets (labelled by
$Z$). On the vertical axis is the
measure given in   of eqn (\ref{quoi1}),
of the relative sensitivity(normalised to Aluminium)  of a target  
to the operators    $O = \widetilde{C}^{pp}_{S,X}$ (continuous black),
 $O = \widetilde{C}^{nn}_{S,X}$(dotted green), $\widetilde{C}^{pp}_{V,X}$ (dashed red) and $O = \widetilde{C}^{nn}_{V,X}$(dot-dashed blue).
%ORDI/ROOT/MEC2/plot4.C 
\label{fig:AlAip2}}
\end{figure}

Perhaps a more transparent measure of the change of direction of 
$\hat{e}_{A}$ in coefficient space, 
is given in  figure \ref{fig:AlAip2} by
the ratio
\beq
  e^O_{A}/e^O_{Al}
\label{quoi1}
\eeq
where  $O = \widetilde{C}^{pp}_{S,X}$ (continuous black),
 $O = \widetilde{C}^{nn}_{S,X}$(dotted green), $\widetilde{C}^{pp}_{V,X}$ (dashed red) and $O = \widetilde{C}^{nn}_{V,X}$(dot-dashed blue).
Recall that $  e^O_{A}$ parametrises the fraction of the sensitivity of  target $A$ to operator $O$. So figure  \ref{fig:AlAip2} shows that heavier targets
 have greater sensitivity to ${\cal O}_V^{nn}$ and   less
 to ${\cal O}_S^{pp}$. (Unfortunately, this figure also suggests
 that  ${\cal O}_V^{nn}$ and  ${\cal O}_S^{pp}$ with comparable
 coefficients  could be difficult to distinguish from
  ${\cal O}_V^{pp}$.)  This normalised ratio of overlap
 integrals is interesting, because the normalisation ``factors out''
 the growth  with $Z$ shared by  all the overlap integrals,
 so this ratio parametrises the  difference in direction in
 coefficient  space, which allows different targets to
discriminate amoung coefficients. 
This ratio also indicates that   targets of $Z \lsim 25$
cannot distinguish  operators, if one admits a
theory uncertainty of $\sim$10\%  in the calculation
of the components  $  e^O_{A}$.

Assisted by the  measures of discriminating power given
in eqns (\ref{ipdiff}) and (\ref{quoi1}), we  now
speculate on a possible series of targets. 
A   light nucleus without spin  could be an interesting
second target, because   it would allow to
distinguish whether the rate in Aluminium was
dominantly SD or SI. In particular,
the SI rate  in Aluminium  could be
approximately  predicted from the
 the rate observed in another  spinless light nucleus.
This is because the SI rate in all targets with
$Z \lsim 20$ is  sensitive to a similar
linear combination
of  operator coefficients, as
 illustrated  in figures \ref{fig:AlAip1}
 and \ref{fig:AlAip2}.

An interesting choice for
the second target could be Titanium (Z=22, A = 48).
As illustrated in figures \ref{fig:AlAip1}
 and \ref{fig:AlAip2}, it of sufficiently
low $Z$  that the SI rate  probes the same combination of  operator
coefficients as the SI rate in Aluminium. So measuring
the SI rate in Titanium-48 would allow to
determine whether there was a significant SD contribution
to  the $\mec$ rate observed  on Aluminium. 

If there is indication for an SD contribution  in
Aluminium, then it could be interesting to measure the
rate on a Titanium target enriched with the spin-carrying
isotopes 47 and 49. This   would give complementary
information on the  quark flavour of the tensor
and/or axial vector operators, because the
spin of Aluminium is largely due to the
odd proton, whereas for Titanium, there
is an odd neutron. So the SD rate in Aluminium
is mostly sensitive to $\widetilde{C}^{'pp}_{A,X}$,
whereas the SD rate in Titanium depends on 
 $\widetilde{C}^{'nn}_{A,X}$.

Finally, if there is no evidence of an
SD rate in Aluminium, a heavy target
such as lead could be interesting 
to discriminate  the scalar vs vector 
coefficients  in  the SI rate.

\section{Summary}
\label{sec:summ}

This  paper gives some details
of the calculation of the  Spin Dependent (SD)
$\mu \to e$ conversion rate  in light nuclei, previously
outlined  in \cite{CDK}. Section \ref{sec:ops}  reviews the operators involving
quarks and gluons that contribute\cite{KKO,CKOT}
at the experimental scale ($\mu_N = 2$ GeV),
and matches them onto the nucleon operators
which enter the nuclear physics calculation.
Some attempt is made to include pion exchange in
this matching (it is relevant because the
momentum-transfer is $m_\mu^2$).
Section \ref{sec:rate}  calculates as much as possible
of the conversion rate in the notation of
relativistic, second-quantised, QFT\cite{P+S};
in the last steps, the results of nuclear
calculations are included. The final rate 
is  given in equation (\ref{BRSD}).
This section is not
original; its purpose is to make the result
accessible to affictionados of QFT.
We recall the SD $\mec$ is incoherent, like
SD WIMP scattering, so it
is best searched for 
in light nuclei,  where the  $1/A^2$
suppression with respect to the coherent Spin
Independent (SI) rat²e (given in eqns (\ref{BRSI},\ref{BRmecKKO}))
is less significant. 

Our  SD rate estimate  relies on  nuclear  physics  calculations
of the   expectation value of nucleon axial currents  in
the nucleus. The results we use were obtained for
SD WIMP scattering, which are often at
zero momentum transfer.  As discussed in
point 6 of section \ref{ssec:CDK}, additional nuclear calculations
seem required  to include tensor and
pseudoscalar operators  at finite momentum transfer,  in 
light targets such as  Aluminium and Titanium.
In this paper, we did not discuss SD conversion
on  heavy nuclei; however, one can speculate
that  the nuclear expectation values could be
of interest, 
%once $\mec$ is observed,
because
heavy nuclei could be sensitive to a different
combination of tensor
and axial operators from light nuclei.
This is because
the anti-lepton wavefunction contributes
with opposite sign to 
the tensor vs  axial operators,
and is  more relevant in heavy nuclei (this sign
difference allows to discriminate scalar
and vector operators in SI conversion on
light and heavy nuclei\cite{KKO}). 
Of course, the SD rate might be unobservably small
(due to the $1/A^2$ suppression), but heavy nuclei
could nonetheless give an independent constraint
on the many operator coefficients.

 Both the SD and SI conversion rates depend on the
modulus-squared of a sum of  coefficients,
weighted by nucleus-dependent numbers--- see equations
(\ref{BRSI},\ref{BRSD},\ref{BRmecKKO}).   This allows for cancellations,
 making it difficult to
constrain individual coefficients,  or identify the
operators responsable for $\mec$ when it is observed. 
In the SI case, Kitano Koike and Okada (KKO)\cite{KKO}  pointed out that 
scalar vs dipole vs vector operators
 could be   distinguished
by changing the nuclear target.
 Section \ref{sec:implications} explores,
 from  various  approaches,
 the prospects of distinguishing  a wider variety of
operators, including   SD vs SI,
and $u$- vs $d$- quark operators.

The prospects for   discriminating  vector or scalar operators
involving either $u$
 or   $d$ quarks  are  illustrated in
figure \ref{fig:uvsd}:
vector operators involving $u$  or $d$ quarks
could be distinguished by comparing the $\mec$ rate in  light
($Z \lsim 20$) and intermediate ($Z \sim 40$) targets,
but distinguishing scalar $u$ versus  $d$ operators seems
difficult. Curiously, the $u$ vs $d$ distinction is
more transparent in the SD rates, as discussed
after eqn (\ref{distinct}). 
So if both SD
and SI conversion  are observed, 
possibly the quark
flavour  could be extracted from the SD rates
\footnote {Recall that SD and SI operators mix in the RG evolution,
but  without changing the quark flavour, as
shown in appendix \ref{app:RGEs}.
The  only flavour change
is via the first two  ``penguin'' diagrams of
figure \ref{fig:diag1}, which could change the flavour of vector
operators.}.

The  SD and SI contributions
to the conversion rate could be distinguished (if the
SD  rate is large enough) by comparing the conversion
rate in nuclei with and without spin.
Section \ref{sec:uncertainties} reviews the theoretical
uncertainties %($\equiv\delta I/I \lsim 10\%$)
in the calculation of the $\mec$ rate, in order to
estimate the sensitivity to the subdominant SD process.
Comparing $\mec$ on  isotopes with and without spin would cancel the leading
theory uncertainties,  giving a sensitivity (see the discussion
after eqn \ref{cancel}) to
 $\Gamma_{SD}/\Gamma_{SI} \gsim \frac{0.1}{A}$, assuming a 10\% uncertainty
 on $\Gamma_{SI}$.
Among the SD operators,  it is not currently  possible to distinguish
pseudoscalar, axial and tensor coefficients, because only the
nuclear expectation value of the axial operator  has been
calculated. However, as mentioned in the previous paragraph, it
 could be possible to discriminate SD operators involving $u$ vs $d$
quarks, because they contribute differently in nuclei where
the odd nucleon is a proton or neutron.

The  upcoming  COMET and Mu2e experiments
will initially search for $\mec$ on 
  Aluminium, a target which has spin ---
so  if  they observe a signal, it could be mediated by the  SD or
SI operators.
So in section  \ref{ssec:EFT}, we considered 
what  series of subsequent  targets could give information
about the dominant coefficients.
To this purpose, we represent 
 a target material  as
a vector in the space of  nucleon-level operators,
whose components are  numbers which multiply
the operator coefficient in the rate (overlap
integrals, in the SI case).  
Different targets  can discriminate between operators,
if they point in different directions of operator
space. We plot in figures \ref{fig:AlAip1}
 and \ref{fig:AlAip2}  two different measures of the
misalignment between target vectors.  

If $\mec$ is observed on Aluminium, 
the following sequence of targets  could be
interesting: as second target, a light nucleus without spin,
such as Titanium-48,
would discriminate whether  the dominant contribution was from
the SD rate, because the SI rate in  Titanium is
comparable to Aluminium (see figures \ref{fig:AlAip1}
 and \ref{fig:AlAip2}).
If there is an SD contribution to the rate
in Aluminium,  then Titanium isotopes with spin, 
could be an interesting target:  the
spin of Titanium  is related to
the odd neutron (whereas in  Aluminium there is an
odd proton), so this could discriminate whether
the SD operators involved $u$ or $d$ quarks.
Finally, a heavy target such as gold or lead could
allow to discriminate  scalar vs vector operators,
as pointed out in \cite{KKO}.

\subsection*{Acknowledgements}

We  greatly thank Vincenzo Cirigliano for his participation in
the early  stages of this project, and for insightful
discussions afterwards. 
SD  acknowledges the partial support and hospitality of
the Mainz Institute for Theoretical Physics (MITP)
during the initial stages  of this work.
The work of Y.K. is
supported in part by the Japan Society of the Promotion
of Science (JSPS) KAKENHI Grant No. 25000004.

\appendix

\section{ $ \bm{G^{N,q}_O}$}
\label{app:A}

When the quark Lagrangian of  eqn (\ref{LVAPST})
is  matched onto the nucleon Lagrangian,
the coefficients of  the nucleon operators
can be computed as 
$\widetilde{C}_{O,Y}^{NN}  = \sum_q G^{N, q}_O C_{O,Y}^{qq}$,
for $O \in T,A,V,P$; for the scalar operator
there  is an  additional gluon contribution as
described in \cite{CKOT}.
We take the $ G^{N,q}_O$, defined  
at zero-momentum-tranfer such that 
$\langle  N(P)| \bar{q}(x) \Gamma_O q(x)|N(P) \rangle$
=$G^{N,q}_O\overline{u_N}(P) \Gamma_O u_N(P)$,
to be
\bea
G^{p,u}_V = G^{n,d}_V = 2 ~~~,~~~ &G^{p,d}_V = G^{n,u}_V = 1& ~~~,~~~
G^{p,s}_V = G^{n,s}_V = 0 
\\
G^{p,u}_A = G^{n,d}_A=  0.84(1) ~~~,~~~ &G^{p,d}_A = G^{n,u}_A = -0.43(1)& ~~~,~~~
G^{p,s}_A = G^{n,s}_A = -.085(18)
\\
G^{p,u}_S = \frac{m_p}{m_u} 0.021(2) =9.0  ~~~,~~~ & G^{p,d}_S = \frac{m_p}{m_d}  0.041(3) = 8.2&    ~~~,~~~  G^{p,s}_S =    \frac{m_N}{m_s}  0.043(11) = 0.42 
\\  G^{n,u}_S = \frac{m_n}{m_u} 0.019(2)= 8.1 ~~~,~~~
&G^{n,d}_S = \frac{m_n}{m_d} 0.045(3)= 9.0&  ~~~,~~~      G^{n,s}_S  \frac{m_N}{m_s}  0.043(11) = 0.42  \\
G^{p,u}_P =144 =G^{n,d}_P   ~~,~~ & G^{p,d}_P = -150  =G^{n,u}_P  &  ~~~,~~
G^{p,s}_P = -4.9  =G^{n,s}_P  \\
G^{p,u}_T = G^{n,d}_T=  0.77(7) ~~~,~~~ &G^{p,d}_T = G^{n,u}_T = -0.23(3)& ~~~,~~~
G^{p,s}_T = G^{n,s}_T = .008(9) ~~~~.
\label{data}
\eea
where the  parenthese gives the uncertainty in the last figure(s).
The axial $G_A$  are  the results inferred in Ref.~\cite{BBPS} 
by using the HERMES measurements~\cite{HERMES}.
The scalar $G_S$ induced by light quarks 
are  from a    dispersive determination~\cite{Hoferichter:2015dsa},
and  an average of lattice results~\cite{Junnarkar:2013ac}
is used for the strange quark; 
in all cases, the 
 $\overline{\rm MS}$  quark masses  at $\mu = 2$~GeV
 are taken as  $m_u = 2.2$~MeV,   $m_d = 4.7$~MeV, and
 $m_s = 96$~MeV~\cite{Agashe:2014kda}.
 The nucleon masses are  $m_p =  938 ~{\rm MeV}$
 and $m_n =  939.6 ~{\rm MeV}$.
The pseudoscalar results  were
calculated from data in  the large-$N_c$ approximation
at  $q^2 \approx 0$  \cite{cheng},
and  here  extrapolated to neutrons using isospin.
The tensor results for
the neutron are  the lattice results of 
Cirigliano etal \cite{ClatticePRL}, which  are here
 extrapolated to protons using isospin.

For comparaison, the $G_A$ have been obtained on the lattice;
a recent determination \cite{Green:2017keo} is 
\bea
G^{p,u}_A = G^{n,d}_A=  0.863(7)(14) ~~~,~~~ &G^{p,d}_A = G^{n,u}_A = -0.345(6)(9)& ~~~,~~~
G^{p,s}_A = G^{n,s}_A = -.0240(21)(11)
\label{GAlattice}
\eea
The scalar $G^{N,q}_S$ have also recently been  obtained on
the lattice \cite{Lellouch}:
\bea
G^{p,u}_S = \frac{m_p}{m_u} 0.0139(13)(12) =5.9 & ~~~,~~~
&G^{p,d}_S = \frac{m_p}{m_d}  0.0253(28)(24) =5.0
\\  G^{n,u}_S = \frac{m_n}{m_u} 0.0116(13)(11)=5.0 & ~~~,~~~
&G^{n,d}_S = \frac{m_n}{m_d} 0.0302(3)=  6.0 
\label{GSlattice}
\eea
We observe  that there is a 50\% discrepancy with
respect to the results of
\cite{Hoferichter:2015dsa}, obtained from pionic atoms and
$\pi-N$ scattering \cite{RuizdeElvira:2017stg}. Results
similar to \cite{Hoferichter:2015dsa} were earlier obtained
in \cite{Alarcon:2011zs}, also using an effective theory.

\section{The  tensor contribution to  the SD and SI rates} \label{app:tensor}

We consider tensor operators
\beq
C^{uu}_{T,L} {\cal O}^{ uu}_{T,L} + C^{dd}_{T,L} {\cal O}^{ dd}_{T,L} + \{ L\leftrightarrow R\}
\eeq
at the experimental scale $\mu_N$, which contribute at
 finite-momentum-transfer to the 
SI conversion process (see eqn (\ref{redef})), and
also to the SD processes:
\bea
\frac{\Gamma_{SI}}{\Gamma_{capt}} &=& 8 B_0\frac{m_\mu^2}{m_N^2}
|Z({C}_{T,L}^{uu}G_T^{p,u}+{C}_{T,L}^{dd}G_T^{p,d})F_p(m_\mu) +
(A-Z) ({C}_{T,L}^{uu}G_T^{n,u}+{C}_{T,L}^{dd}G_T^{n,d})F_n(m_\mu)|^2 + \{L \leftrightarrow R\}
\label{BRSIt}
\\
\frac{\Gamma_{SD}}{\Gamma_{capt}}
&=& 32 B_0\frac{J_{A}+1}{J_{A}}
\, \Big|   S^{A}_p ({C}_{T,L}^{uu}G_T^{p,u}+{C}_{T,L}^{dd}G_T^{p,d})  + 
 S^{A}_n ({C}_{T,L}^{uu}G_T^{n,u}+{C}_{T,L}^{dd}G_T^{n,d}) \Big|^2    \  \frac{S_{A} (m_\mu)}{S_{A} (0)} 
+ \{ L \leftrightarrow R \}~.
\label{BRSDt}
\eea
The ratio of these  contributions, for a single operator, is
\bea
\frac{\Gamma_{SD}}{\Gamma_{SI}} \simeq 4 \frac{J_A +1}{J_A} \frac{m_N^2}{m_\mu^2}  \frac{
\Big|   S^{A}_p G_T^{p,q} + 
 S^{A}_n G_T^{n,q} \Big|^2 }{|Z G_T^{p,q} +
(A-Z)G_T^{n,q}|^2} \sim
\left\{
\begin{array}{lll}
0.7 &q=u&A = {\rm Al}\\
0.06 &q=d&A = {\rm Al}\\
0.03 &q=u&A = {\rm Ti}\\
0.01  &q=d&A = {\rm Ti}\\
\end{array}
\right.
\label{estimates}
\eea
where we assumed  that the form factors are comparable
$\frac{S_{A} (m_\mu)}{S_{A} (0)}  \simeq |F_p(m_\mu)|^2$ as
is the case in Aluminium.  Recall that $G_T^{n,u} \sim -\frac{1}{2}G_T^{p,u}$,
so there is a partial cancellation in the SI amplitude, whereas
the SD process 
arises mostly from an odd proton
$ S^{A}_p \gg S^{A}_n$, or mostly from an odd neutron 
$ S^{A}_p \ll S^{A}_n$.

The estimates  of eqn (\ref{estimates}) assume that
only one tensor coefficient is non-zero, so they 
neglect interferences, which can easily enhance the
SI rate. For instance,  RG running of the tensor operator
from the New Physics scale to the experimental scale 
 generically generates a scalar operator with comparable
 coefficient. The  scalar-tensor interference
 contribution to the SI rate would be relatively enhanced,
 with respect to the tensor-squared, by $G_S^{N,q}/G_T^{N,q}\sim 10$,
 which would suppress the ratio in eqn (\ref{estimates})
 by another factor $1/10$.

\section{RG Evolution} \label{app:RGEs}

In this appendix, we  review the  Renormalisation Group evolution
of operator coefficients from
the leptoquark mass scale $M$ ($\sim$ TeV) down to the
experimental scale $\mu_{N}$ (2 GeV), via the one-loop
RGEs of  QCD and QED \cite{PSI,megmW}. We consider the
QED$\times$ QCD  invariant operator basis discussed
in section \ref{sec:ops}. We  neglect matching onto
the SMEFT basis \cite{BW,polonais} and running with
the full SM RGEs\cite{JMT}, on the assumption that QED
is a reasonable approximation  if $M$ is not much larger than $m_W$.

After including one-loop corrections in the
$\overline{MS}$ scheme,  the operator
coefficients will run with scale $\mu$ according to\cite{megmW}
\begin{equation}
\mu \frac{\partial}{\partial \mu} (C_{I},...C_{J},...)= \frac{\alpha_{e}} {4\pi}\overrightarrow{C} \Gamma^{e} + \frac{\alpha_{s}} {4\pi} \overrightarrow{C} \Gamma^{s}
\end{equation}
where $I,J$ represent the super- and subscripts which label
operator coefficients, $\Gamma^{e}$ and $\Gamma^{s}$ are the
QED and QCD anomalous dimension matrices and
$\overrightarrow{C}$  is a  row vector that
contains the QCD $\times$QED invariant operators coefficients listed in section 2. 

In this work, we use the approximate analytic solution\cite{CDK}
given in eqn (\ref{oprun1l}):    
\begin{equation} \nonumber
 C_{I} (\mu_{N})  = C_{J}(M)\lambda^{a_{J}}
\left( \delta_{JI} - \frac{\alpha_{e}  \widetilde{\Gamma}^{e}_{JI} }{4\pi}  \log \frac{M}{\mu_{N}}  \right)
\label{RGE}
\end{equation}
 where the factors are given after
eqn (\ref{oprun1l}) and $\log \frac{M}{\mu _{N}} \sim$ 6.21.

\begin{figure}[h]
\begin{center}
\epsfig{file=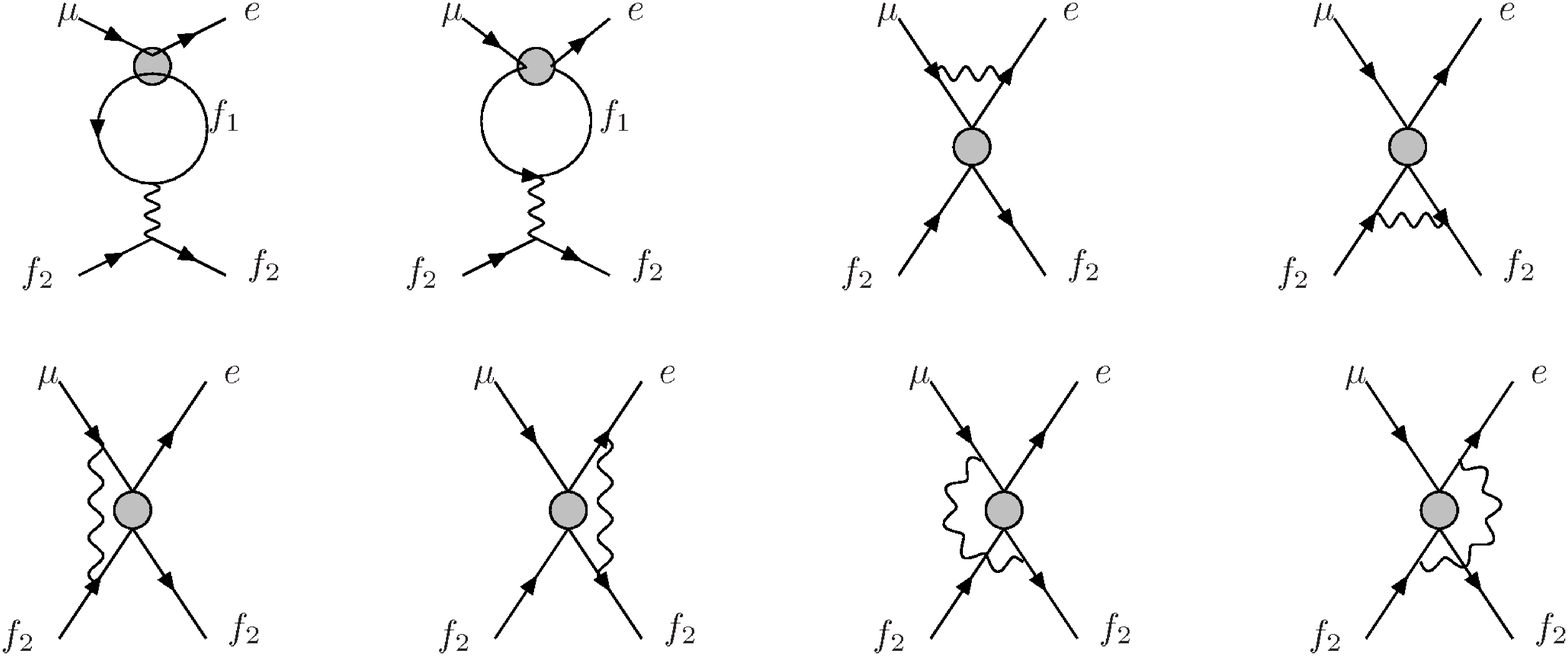,height=7cm,width=16cm}
\end{center}
\caption{ Examples of one-loop gauge vertex corrections to 4-fermion operators.
The first two  diagrams are the penguins.
The last six diagrams  contribute to operator
mixing and running, but can  only change the Lorentz or gauge structure
of the operators, not the flavour structure.
Missing are the wave-function renormalisation diagrams; for $V \pm A$
Lorentz structure in the grey blob, this cancels diagrams 3 and 4.
 \label{fig:diag1}
}
\end{figure}

 Only QED loops  contribute to operators mixing, while QCD
 loops only rescale scalar and tensor operators.
 In figure \ref{fig:diag1}, we present the QED diagrams
 required to compute the anomalous dimension $\gamma$ of
 the four-fermion operators, where
 $f_{1} \in \left\{e, \mu \right\}$
 and $f_{2} \in \left\{u,d,s,c,b,e, \mu, \tau \right\}$.   

 The operators coefficients below the scale $M$ are organized in the   vector $\overrightarrow{C}$ as following  :

\begin{align}
&\vec{C} = (\vec{C}^{u}_V,  \vec{C}^{d}_V,\vec{C}^{u}_A,  \vec{C}^{d}_A, \vec{C}^{u}_S,  \vec{C}^{d}_S,    \vec{C}^{u}_T,    \vec{C}^{d}_T)\\ 
&\vec{C}_V^{ f}  = ( C_{VL}^{f f} , C_{VR}^{ f f}) ~~~ \vec{C}_A^{ f}  = ( C_{AL}^{f f} , C_{AR}^{ f f}) \\
&\vec{C}_S^{f} = ( C_{S,L}^{f f} , C_{S,R}^{f f}) ~~~ \vec{C}_T^{f}  = ( C_{T,L}^{ff}  , C_{T,R}^{ ff} ) \\
\end{align}
 In the basis of $\overrightarrow{C}$,  the QED anomalous dimension matrix can be written 
$$\Gamma^{e}=
\begin{bmatrix} 
\Gamma_{VA} & 0 \\
0 & \Gamma_{ST} 
\end{bmatrix}
$$ where
\begin{align}
&\Gamma_{ST}=
\left[
\begin{array}{cccccccccccc} 
& \gamma_{S,S}^{u,u} &0   & \gamma_{S,T}^{u,u} &0  \\
&0 & \gamma_{S,S}^{d,d}   &0 & \gamma_{S,T}^{d,d}  \\
&\gamma_{T,S}^{u,u}     &0 &\gamma_{T,T}^{u,u}&0    \\
&0  &\gamma_{T,S}^{d,d}   &0 &\gamma_{T,T}^{d,d}     \\
\end{array}
\right] 
& {\rm and} ~~~~~~ \Gamma_{VA}=
\left[
\begin{array}{cccccccccccc} 
& 0 &0   & \gamma_{V,A}^{u,u} &0  \\
&0 & 0   &0 & \gamma_{V,A}^{d,d}  \\
&\gamma_{A,V}^{u,u}     &0 &0  &0   \\
&0  &\gamma_{A,V}^{d,d}   &0 &0     \\
\end{array}
\right]
\end{align}

\noindent  \textbf{Vector and axial operators} \\

 %%%%%%%%%%%%%%%%%%%%%%%%%%%%%%%%%%%%%%%%%%%%%%%%%%%%%%%%
\iffalse
\begin{equation}
[\gamma_{V,V}^{f_1,f_2}] =
% = \left[
\begin{array}{c|cccc}
 &C^{LL}_{ f_2 f_2} & C^{RR}_{f_2 f_2} & C^{LR}_{ f_2 f_2} 
& C^{RL}_{ f_2 f_2}\\
\hline 
C^{LL}_ { f_1 f_1} & \frac{4}{3}Q_{f_1} Q_{f_2} N_{c,f} &0&
 \frac{4}{3}Q_{f_1} Q_{f_2} N_{c,f} &0\\
C^{RR}_{ f_1 f_1} &  0& 
\frac{4}{3}Q_{f_1} Q_{f_2} N_{c,f} &0&
 \frac{4}{3}Q_{f_1} Q_{f_2} N_{c,f} \\
C^{LR}_{ f_1 f_1} &  
\frac{4}{3}Q_{f_1} Q_{f_2} N_{c,f} &0&
 \frac{4}{3}Q_{f_1} Q_{f_2} N_{c,f} &0\\
C^{RL}_{ f_1 f_1} & 0& 
\frac{4}{3}Q_{f_1} Q_{f_2} N_{c,f} &0&
 \frac{4}{3}Q_{f_1} Q_{f_2} N_{c,f} \\
\end{array}
\end{equation}

For $f_1 \neq  f_2 $ and $f_1 \in \{ e,\mu \}$, the two
penguin diagrams give
\begin{equation}
[g_{V,V}^{f_1,f_2}] =
\left[
\begin{array}{cccc}
-\frac{8}{3} Q_{f_2}  &0&
 -\frac{8}{3} Q_{f_2}  &0\\
  0& 
-\frac{8}{3} Q_{f_2}  &0&
 -\frac{8}{3} Q_{f_2}  \\
  -\frac{4}{3} Q_{f_2} &0&
 -\frac{4}{3} Q_{f_2}  &0\\
 0& 
-\frac{4}{3} Q_{f_2}  &0&
 -\frac{4}{3} Q_{f_2}  \\
\end{array}
\right]
\end{equation}
\fi
 %%%%%%%%%%%%%%%%%%%%%%%%%%%%%%%%%%%%%%%%%%%%%%%%%%%%%%%%

\noindent The first  penguin diagram and the last four give the following matrices :
\begin{equation}
 \gamma_{V,A}^{f,f} =
\begin{array}{c|cc}
&C_{A,L}^{f f} & C_{A, R}^{f f} \\ \hline
C_{V,L}^{ f f}& -12 Q_{f} &0 \\
C_{V,R}^{f f}&0& -12 Q_{f}  \\
\end{array} ~~~
 \gamma_{A,V}^{f,f} =
\begin{array}{c|cc}
&C_{V,L}^{f f} & C_{V, R}^{f f} \\ \hline
C_{A,L}^{ f f}& 12 Q_{f} &0 \\
C_{A,R}^{f f}&0& 12 Q_{f}  \\
\end{array}
%\right]
\end{equation}

%%%%%%%%%%%
\iffalse
\begin{equation}
\gamma_{V,V}^{f,f} =
\left[
\begin{array}{cccc}
-12 Q_{f}+\frac{4}{3}Q_{f}^2  N_{c,f} &0&
 \frac{4}{3}Q_{f}^2  N_{c,f} &0\\
  0& -12 Q_{f}+\frac{4}{3}Q^2_{f}  N_{c,f} &0&
 \frac{4}{3}Q_{f} ^2 N_{c,f} \\
  \frac{4}{3}Q_{f} ^2 N_{c,f} &0&
 12 Q_{f}+\frac{4}{3}Q_{f} ^2 N_{c,f} &0\\
 0& \frac{4}{3}Q_{f}^2 N_{c,f} &0&
 12 Q_{f} +\frac{4}{3}Q_{f} ^2 N_{c,f} \\
\end{array}
\right]
\label{gammaVV}
\end{equation}
\fi
%%%%%%%%%%%%%%%%%%%%

\noindent Using these anomalous dimension matrices and the RGEs give  : 

\begin{align}
&C_{V,R}^{qq}(\mu_{N}) = -3Q_{q}\frac{\alpha_{e}}{\pi} \log \frac{M}{\mu_{N}}C_{A,L}^{qq}(M) + C_{V,R}^{qq}(M) \\
&C_{A,R}^{qq}(\mu_{N}) = 3Q_{q}\frac{\alpha_{e}}{\pi} \log \frac{M}{\mu_{N}}C_{V,L}^{qq}(M) + C_{A,R}^{qq}(M)
\label{oprun1la}
\end{align}
\noindent where $q \in \left\{u,d \right\}$. We see that  axial operators mix to vector operators and vice versa, but there is no rescaling for  axial and vector operators. \\

\noindent  \textbf{Scalar operators} \\

\noindent Combining the third and fourth diagrams of figure \ref{fig:diag1} with the wavefunction diagrams renormalize the scalars  while the last four diagrams mix the tensors  to the scalars :

\begin{equation}
 \gamma_{S,S}^{f,f} =
\begin{array}{c|cc}
&C_{S,L}^{f f} & C_{S, R}^{f f} \\ \hline
C_{S,L}^{ f f}& 6(1 + Q_f^2) &0 \\
C_{S,R}^{f f}&0& 6(1 + Q_f^2)  \\
\end{array} ~~~
\gamma_{T,S}^{f,f} =
\begin{array}{c|cc}
&C_{S, L}^{f f} & C_{S, R}^{f f} \\ \hline
C_{T, L}^{ f f}& -96 Q_f &0 \\
C_{T,R}^{f f}&0& -96 Q_f  \\
\end{array}
%\right]
\end{equation}

\noindent The scalars coefficients at the experimental scale read : 

\begin{equation}
C_{S,L}^{qq}(\mu_{N}) =  24 \lambda^{a_{T}}f_{TS}Q_{q}\frac{\alpha_{e}}{\pi} \log \frac{M}{\mu_{N}}C_{T,L}^{qq}(M) +
\lambda^{a_{S}}\left[1-\frac{3}{2}  \frac{\alpha_{e}  }{\pi}\log \frac{M}{\mu_{N}}(1+Q_{q}^{2})\right]C_{S,L}^{qq}(M) 
\label{oprun1lb}
\end{equation}

\noindent \textbf{Tensor operators} \\

\noindent Similarly, the last four diagrams mix  the scalars  to the  tensors. Only the wavefunction diagrams renormalize the tensors, because for the third and fourth diagrams $\gamma^{\mu} \sigma \gamma_{\mu}=0$. We obtain the following matrices :

\begin{equation}
 \gamma_{T,T}^{f,f} =
\begin{array}{c|cc}
&C_{T,L}^{f f} & C_{T, R}^{ f f} \\ \hline
C_{T,L}^{ f f}& -2(1 + Q_f^2) &0 \\
C_{T,R}^{ f f}&0& -2(1 + Q_f^2)  \\
\end{array}
\hspace{1cm}
\gamma_{S,T}^{f,f} = 
%\left[
\begin{array}{c|cc}
&C_{T,L}^{ f f} & C_{T, R}^{ f f} \\ \hline
C_{S,L}^{ f f}& 2 Q_f &0 \\
C_{S,R}^{ f f}&0& 2Q_f \\
\end{array}
\end{equation}

\begin{equation}
C_{T,L}^{qq}(\mu_{N}) =  - \lambda^{a_{S}}f_{ST}Q_{q}\frac{\alpha_{e}}{2\pi} \log \frac{M}{\mu_{N}}C_{S,L}^{qq}(M) +
 \lambda^{a_{T}}\left[1+\frac{\alpha_{e}}{2\pi}\log \frac{M}{\mu_{N}}(1+Q_{q}^{2})\right]C_{T,L}^{qq}(M)
\label{oprun1lc}
\end{equation}

%%%%%%%%%%%%%%%%%%%
\iffalse
\noindent \textbf{Dipole operators} \\

\noindent The first penguin diagram of figure \ref{fig:diag1} (with the $f_{2}$ line removed) mixes the tensors to the dipoles while the  second penguin diagram mixes the scalars 
to the dipoles :

\begin{equation}
 \gamma_{S,D}^{ff} =
%\left[
\begin{array}{c|cc}
&C^{e \mu}_{D,L} & C^{e \mu}_{D,R}
\\ \hline
C_{S,LL}^{e \mu f f }& -\frac{m_f }{e m_\mu} &0\\
C_{S,RR}^{e \mu f f }&0& -\frac{m_f }{e m_\mu} \\
\end{array}
%\right]
\hspace{1cm}
 \gamma_{T,D}^{ff} =
\begin{array}{c|cc}
&C^{e \mu}_{D,L} & C^{e \mu}_{D,R} \\ \hline
C_{T,LL}^{e \mu ff}& 8 \frac{Q_fN_c m_f}{m_\mu e} &0 \\
C_{T,RR}^{e \mu ff}&0& 8 \frac{Q_fN_c m_f}{ m_\mu e}  \\
\end{array}
\end{equation}

\noindent where f $\in \left\{ e, \mu\right\}$. The dipole operators are now written 

\begin{equation}
C_{D,L}^{e\mu}(\mu_{N}) = -\frac{2Q_{u}N_{c}m_{u}}{e m_{\mu}}\frac{\alpha_{e}}{\pi} \log \frac{M}{\mu_{N}} C_{T,L}^{uu}(M) + \frac{m_{u}}{4e m_{\mu}}\frac{\alpha_{e}}{\pi} \log \frac{M}{\mu_{N}} C_{S,L}^{uu}(M)
\label{oprun1le}
\end{equation} \\

We note that dimension six  ($O_{S,X}^{qq},O_{T,X}^{qq},O_{V,X}^{qq}$ or $O_{A,X}^{qq}$) and dimension five operators ($O_{D,X}$) do not mix to the operators $O_{GG,X}$ which are of dimension seven.
\fi
%%%%%%%%%%ùù

\noindent Finally, the coefficients at the experimental scale $\mu_{N}$ are obtain via the matching condition : 

\begin{equation}
\tilde{C}^{NN}_{O,Y}(\mu_{N}) = \sum_{q=u,d,s}    G_O^{N,q}   C_{O,Y}^{qq}(\mu_{N}) \\
\end{equation}

\end{document}